\documentstyle[12pt,epsf,rotate]{article}

\def\spose#1{\hbox to 0pt{#1\hss}}
\def\lsim{\mathrel{\spose{\lower 3pt\hbox{$\mathchar"218$}}
 \raise 2.0pt\hbox{$\mathchar"13C$}}}
\def\gsim{\mathrel{\spose{\lower 3pt\hbox{$\mathchar"218$}}
 \raise 2.0pt\hbox{$\mathchar"13E$}}}

\begin{document}

\begin{titlepage}

\begin{flushright}
CERN-TH/98-319\\
TUM-HEP-330/98\\
hep-ph/9810260 
\end{flushright}

\vspace{0.7cm}
\begin{center}
\boldmath
\large\bf
A General Analysis of $\gamma$ Determinations from $B\to\pi K$ Decays 
\unboldmath
\end{center}

\vspace{0.6cm}
\begin{center}
Andrzej J. Buras$^{1,2}$ \,and\, Robert Fleischer$^{1}$\\[0.5cm]
{\sl $^{1}$Theory Division, CERN, CH-1211 Geneva 23, Switzerland}\\[0.3cm]
{\sl $^{2}$Technische Universit\"at M\"unchen, Physik Department\\ 
D-85748 Garching, Germany}\\[0.6cm]
\end{center}

\vspace{0.7cm}
\begin{abstract}
\vspace{0.2cm}\noindent
We present a general parametrization of $B^\pm\to\pi^\pm K$, $\pi^0K^\pm$ 
and $B_d\to\pi^0 K$, $\pi^\mp K^\pm$ decays, taking into account both 
electroweak penguin and rescattering effects. This formalism allows -- among
other things -- an improved implementation of the strategies that were 
recently proposed by Neubert and Rosner to probe the CKM angle $\gamma$ with 
the help of $B^\pm\to\pi^\pm K$, $\pi^0K^\pm$ decays. In particular, it allows 
us to investigate the sensitivity of the extracted value of $\gamma$ to the 
basic assumptions of their approach. We find that certain $SU(3)$-breaking 
effects may have an important impact and emphasize that additional hadronic 
uncertainties are due to rescattering processes. The latter can be controlled 
by using $SU(3)$ flavour symmetry arguments and additional experimental 
information provided by $B^\pm\to K^\pm K$ modes. We propose a new strategy 
to probe the angle $\gamma$ with the help of the neutral decays 
$B_d\to\pi^0 K$, $\pi^\mp K^\pm$, which is theoretically cleaner than the 
$B^\pm\to\pi^\pm K$, $\pi^0K^\pm$ approach. Here rescattering processes 
can be taken into account by just measuring the CP-violating observables 
of the decay $B_d\to\pi^0K_{\rm S}$. Finally, we point out that an 
experimental analysis of $B_s\to K^+K^-$ modes would also be very 
useful to probe the CKM angle $\gamma$, as well as electroweak penguins, 
and we critically compare the virtues and weaknesses of the various approaches 
discussed in this paper. As a by-product, we point out a strategy to 
include the electroweak penguins in the determination of the CKM angle
$\alpha$ from $B\to\pi\pi$ decays. 
\end{abstract}

\vfill
\noindent
CERN-TH/98-319\\
October 1998

\end{titlepage}

\thispagestyle{empty}
\vbox{}
\newpage
 
\setcounter{page}{1}

\section{Introduction}\label{intro}
Last year, the CLEO collaboration reported the observation of several 
exclusive $B$-meson decays into two light pseudoscalar mesons \cite{cleo},
which led to great excitement in the $B$-physics community. In particular,
the decays $B^+\to\pi^+K^0$, $B^0_d\to\pi^-K^+$ and their charge conjugates
received a lot of attention \cite{ICHEP98}, since their observables may 
provide useful information on the angle $\gamma$ of the usual 
non-squashed unitarity triangle of the Cabibbo--Kobayashi--Maskawa matrix 
(CKM matrix) \cite{PAPIII,groro}. So far, only results for the combined 
branching ratios
\begin{eqnarray}
\mbox{BR}(B^\pm\to\pi^\pm K)&\equiv&\frac{1}{2}\left[\mbox{BR}(B^+\to\pi^+K^0)
+\mbox{BR}(B^-\to\pi^-\overline{K^0})\right]\\
\mbox{BR}(B_d\to\pi^\mp K^\pm)&\equiv&\frac{1}{2}\left[\mbox{BR}(B^0_d
\to\pi^-K^+)+\mbox{BR}(\overline{B^0_d}\to\pi^+K^-)\right]
\end{eqnarray}
have been published, with values at the $10^{-5}$ level and large experimental 
uncertainties. As was pointed out in \cite{fm2}, already these combined
branching ratios may lead to highly non-trivial constraints on $\gamma$,
which become effective if the ratio
\begin{equation}\label{Def-R}
R\equiv\frac{\mbox{BR}(B_d\to\pi^\mp K^\pm)}{\mbox{BR}(B^\pm\to\pi^\pm K)}
\end{equation}
is found to be smaller than 1. If we use the $SU(2)$ isospin symmetry of
strong interactions and neglect certain rescattering and electroweak penguin 
effects (for more sophisticated strategies, taking into account also these 
effects, see \cite{defan,rf-FSI}), we obtain the following allowed range 
for $\gamma$ \cite{fm2}:
\begin{equation}\label{gamma-bound1}
0^\circ\leq\gamma\leq\gamma_0\quad\lor\quad180^\circ-\gamma_0\leq\gamma
\leq180^\circ,
\end{equation}
where $\gamma_0$ is given by
\begin{equation}\label{gam0}
\gamma_0=\arccos(\sqrt{1-R})\,.
\end{equation} 
Unfortunately, the present data do not yet provide a definite answer to the 
question of whether $R<1$. The results reported by the CLEO collaboration 
last year gave $R=0.65\pm0.40$ \cite{cleo}, whereas a recent, preliminary
update yields $R=1.0\pm0.4$ \cite{newCLEO}. A detailed study of the 
implications of (\ref{gamma-bound1}) for the determination of the unitarity 
triangle was performed in \cite{gnps}.

This summer, the CLEO collaboration announced the first observation
of another $B\to\pi K$ decay, which is the mode $B^\pm\to\pi^0K^\pm$ 
\cite{newCLEO}. Consequently, it is natural to ask whether we could 
also obtain interesting information on the angle $\gamma$ with the help of
this decay. In fact, several years ago, Gronau, Rosner and London (GRL) 
proposed an interesting strategy to determine $\gamma$, with the help of the 
decays $B^+\to\pi^0K^+$, $B^+\to\pi^+K^0$, $B^+\to\pi^+\pi^0$ and their charge 
conjugates, by using the $SU(3)$ flavour symmetry of strong interactions 
\cite{grl} (see also \cite{hlgr}). However, as was pointed out by Deshpande 
and He \cite{dh}, this elegant approach is unfortunately spoiled by 
electroweak penguins, which play an important role in several non-leptonic 
$B$-meson decays because of the large top-quark mass 
\cite{rf-ewp,rf-ewp3}. In the case of the mode
$B^+\to\pi^0K^+$, electroweak penguins contribute both in ``colour-allowed'' 
and in ``colour-suppressed'' form, whereas only electroweak penguin
topologies of the latter kind contribute to the decays $B^+\to\pi^+K^0$ and 
$B^0_d\to\pi^-K^+$. Performing model calculations within the framework of 
the ``factorization'' hypothesis, one finds that ``colour-suppressed''
electroweak penguins play a negligible role \cite{akl}. These crude 
estimates may, however, underestimate the role of these topologies
\cite{groro,neubert}, which therefore represent an important limitation of 
the theoretical accuracy of the strategies to probe the CKM angle $\gamma$ 
with the help of $B^\pm\to\pi^\pm K$ and $B_d\to\pi^\mp K^\pm$ decays 
\cite{ICHEP98}.

In \cite{PAPIII,PAPI}, we proposed methods to obtain experimental insights 
into electroweak penguins with the help of amplitude relations between the 
$B\to\pi K$ decays listed above. Since it is possible to derive a transparent
expression for the relevant electroweak penguin amplitude by performing
appropriate Fierz transformations of the electroweak penguin operators and 
using the $SU(3)$ flavour symmetry \cite{PAPIII} (see also \cite{rf-ewp3}), 
the experimental determination of this amplitude would allow an interesting 
test of the Standard Model. In two recent papers \cite{nr1,nr2}, Neubert and 
Rosner used a more elaborate, but similar theoretical input to calculate the 
electroweak penguin amplitude affecting the GRL approach. Provided the 
electroweak penguin amplitude calculated this way is theoretically 
reliable, the combined $B^\pm\to\pi^\pm K$ and $B^\pm\to\pi^0 K^\pm$ branching 
ratios may imply interesting bounds on the CKM angle $\gamma$ \cite{nr1}, 
and the original GRL strategy, requiring the measurement of a CP-violating 
asymmetry in $B^\pm\to\pi^0 K^\pm$, is resurrected \cite{nr2}.

In this paper, we point out that the general formulae to probe the CKM angle 
$\gamma$, with the help of the decays $B^\pm\to\pi^\pm K$ and 
$B_d\to\pi^\mp K^\pm$ that were derived in \cite{defan}, apply also to the 
combination $B^\pm\to\pi^\pm K$, $\pi^0 K^\pm$ of charged $B$ decays, 
as well as to the combination $B_d\to\pi^0K$, $\pi^\mp K^\pm$ of neutral
$B$ decays, if straightforward replacements of variables are performed. In 
this manner, the virtues and weaknesses of the strategies proposed in
\mbox{\cite{defan,nr1,nr2}}, and of a new one proposed here, can be 
systematically investigated and compared with one another. Following these 
lines, we are in a position to derive the bounds on $\gamma$ arising in the 
$B^\pm\to\pi^\pm K$, $\pi^0 K^\pm$ case \cite{nr1} in a general and 
transparent way. In contrast to the expressions given in \cite{nr1}, our 
formalism is valid {\it exactly} and does not rely on any expansion in a 
small parameter. Moreover, it allows us to investigate the sensitivity of the 
value of $\gamma$ to the basic assumptions made in \cite{nr1}, and to take 
into account certain rescattering processes by using the strategies proposed 
in \cite{defan,rf-FSI}. These final-state interaction effects were neglected 
by Neubert and Rosner in \cite{nr1,nr2}, but may in principle play an 
important role [16, 20--24]. We find that 
they lead to uncertainties similar to those affecting the $B^\pm\to\pi^\pm K$, 
$B_d\to\pi^\mp K^\pm$ strategies \cite{ICHEP98}, and that furthermore 
certain $SU(3)$-breaking effects may have an important impact. Concerning 
rescattering effects, the neutral decays $B_d\to\pi^0K$, 
$\pi^\mp K^\pm$ offer theoretically cleaner strategies to probe the 
CKM angle $\gamma$ than the charged modes $B^\pm\to\pi^\pm K$, $\pi^0 K^\pm$. 
The point is that the decay $B_d\to\pi^0K_{\rm S}$ provides an additional 
observable, which originates from mixing-induced CP violation. If we use 
in addition the CP asymmetry arising in the mode $B_d\to J/\psi\, K_{\rm S}$ 
to fix the $B^0_d$--$\overline{B^0_d}$ mixing phase, the rescattering 
processes can be included completely. We also point out that an 
experimental analysis of the decay $B_s\to K^+K^-$ would offer -- in 
combination with the data provided by $B_d\to\pi^\mp K^\pm$, 
$B^\pm\to\pi^\pm K$ and $B^\pm\to\pi^\pm\pi$ -- several simple strategies 
both to probe the CKM angle $\gamma$ and to obtain insights into electroweak 
penguins. Therefore, an accurate measurement of the decay $B_s\to K^+K^-$, 
which should be feasible at future hadron machines, would be an important 
goal. 

The outline of this paper is as follows: in Section~\ref{DAO}, we present
a general parametrization of the $B\to\pi K$ decay amplitudes and observables, 
taking into account both electroweak penguin and rescattering effects.
In Section~\ref{PAPIII-strat}, we recapitulate the $B^\pm\to\pi^\pm K$, 
$B_d\to\pi^\mp K^\pm$ strategies to constrain and determine the CKM angle 
$\gamma$ in the light of the most recent CLEO data, and point out some 
interesting features that were not emphasized in previous work. In 
Section~\ref{char-strat}, we focus on strategies to probe $\gamma$ with the 
help of the charged decays $B^\pm\to\pi^\pm K$, $\pi^0 K^\pm$, while we 
turn to a new approach, using the neutral modes $B_d\to\pi^0K$, 
$\pi^\mp K^\pm$, in Section~\ref{neut-strat}. Several strategies to combine 
the observables of the $B_{u,d}\to\pi K$ modes with those of the decay 
$B_s\to K^+K^-$ to determine the CKM angle $\gamma$ and to probe 
electroweak penguins are proposed in Section~\ref{BsKK}. Finally, the 
conclusions are summarized in Section~\ref{concl}, where we also critically
compare the virtues and weaknesses of the various approaches discussed 
in this paper. In an appendix, we present a by-product of our considerations, 
allowing us to include electroweak penguin topologies in the determination 
of the CKM angle $\alpha$ from $B\to\pi\pi$ decays.

\section{Decay Amplitudes and Observables}\label{DAO}
In this section, we will closely follow Ref.\ \cite{defan} to parametrize 
the decay amplitudes and observables of $B^\pm\to\pi^0 K^\pm$
and $B_d\to\pi^0K$ arising within the framework of the Standard Model. 
Before turning to these modes, it will be instructive to recall certain
features of the decays $B^\pm\to\pi^\pm K$ and $B_d\to\pi^\mp K^\pm$, 
which were already discussed in detail in \cite{defan}. 

\boldmath
\subsection{The Decays $B^\pm\to\pi^\pm K$ and 
$B_d\to\pi^\mp K^\pm$}
\unboldmath
In order to probe the CKM angle $\gamma$ through these decays,
the central role is played by the following amplitude relation:
\begin{equation}\label{ampl-rel1}
A(B^+\to\pi^+K^0)+A(B_d^0\to\pi^-K^+)=-\left[T+P_{\rm ew}^{\rm C}\right],
\end{equation}
which can be derived by using the $SU(2)$ isospin symmetry of 
strong interactions \cite{bfm}. Here the amplitude $T$, which is usually
referred to as a ``tree'' amplitude, takes the form 
\begin{equation}\label{T-def}
T=|T|e^{i\delta_T}e^{i\gamma}. 
\end{equation}
Owing to a subtlety in the implementation of the isospin symmetry, the 
amplitude $T$ does not only receive contributions from colour-allowed 
$\bar b\to\bar uu\bar s$ tree-diagram-like topologies, but also from 
penguin and annihilation topologies \cite{defan,bfm}. On the other hand,
the quantity $P_{\rm ew}^{\rm C}$ is due to electroweak penguin contributions,
which do not carry the phase $e^{i\gamma}$, and can be expressed as 
\begin{equation}\label{PewC-def}
P_{\rm ew}^{\rm C}=-\,|P_{\rm ew}^{\rm C}|e^{i\delta_{\rm ew}^{\rm C}}.
\end{equation}
Note that the remaining electroweak penguin contributions have been
absorbed in the amplitude $T$. The label ``C'' reminds us that only 
``colour-suppressed'' electroweak penguin topologies contribute to 
$P_{\rm ew}^{\rm C}$. In (\ref{T-def}) and (\ref{PewC-def}), 
$\delta_T$ and $\delta_{\rm ew}^{\rm C}$ denote CP-conserving strong phases. 
Explicit formulae for $T$ and $P_{\rm ew}^{\rm C}$ are given in \cite{defan}.

The $B^+\to\pi^+K^0$ decay amplitude entering (\ref{ampl-rel1}) can be 
expressed as follows \cite{defan}:
\begin{equation}
A(B^+\to\pi^+K^0)=\lambda^{(s)}_u\left[P_u+P_{\rm ew}^{(u){\rm C}}+
{\cal A}\right]+\lambda^{(s)}_c\left[P_c+P_{\rm ew}^{(c){\rm C}}\right]+
\lambda^{(s)}_t\left[P_t+P_{\rm ew}^{(t){\rm C}}\right],
\end{equation}
where $P_q$ and $P_{\rm ew}^{(q){\rm C}}$ denote contributions from QCD 
and electroweak penguin topologies with internal $q$ quarks $(q\in\{u,c,t\})$,
respectively; ${\cal A}$ describes annihilation topologies, and 
$\lambda^{(s)}_q\equiv V_{qs}V_{qb}^\ast$ are the usual CKM factors.
If we make use of the unitarity of the CKM matrix and apply the Wolfenstein 
parametrization \cite{wolf}, generalized to include non-leading terms
in $\lambda$ \cite{blo}, we obtain \cite{defan}
\begin{equation}\label{Bpampl}
A(B^+\to\pi^+K^0)\equiv P=-\left(1-\frac{\lambda^2}{2}\right)\lambda^2A
\left[1+\rho\,e^{i\theta}e^{i\gamma}\right]{\cal P}_{tc}\,,
\end{equation}
where
\begin{equation}\label{Ptc}
{\cal P}_{tc}\equiv\left|{\cal P}_{tc}\right|e^{i\delta_{tc}}=
P_t-P_c+P_{\rm ew}^{(t){\rm C}}-P_{\rm ew}^{(c){\rm C}}
\end{equation}
and
\begin{equation}\label{rho-def}
\rho\,e^{i\theta}=\frac{\lambda^2R_b}{1-\lambda^2/2}
\left[1-\left(\frac{{\cal P}_{uc}+{\cal A}}{{\cal P}_{tc}}\right)\right].
\end{equation}
In these expressions, $\delta_{tc}$ and $\theta$ denote CP-conserving 
strong phases, and ${\cal P}_{uc}$ is defined in analogy to  (\ref{Ptc}). 
The quantity $\rho\,e^{i\theta}$ is a measure of the strength of certain 
rescattering effects, and the relevant CKM factors are given by (for a 
recent update of $R_b$, see \cite{Rb-update}):
\begin{equation}\label{CKM-exp}
\lambda\equiv|V_{us}|=0.22\,,\quad A\equiv\frac{1}{\lambda^2}
\left|V_{cb}\right|=0.81\pm0.06\,,\quad R_b\equiv\frac{1}{\lambda}
\left|\frac{V_{ub}}{V_{cb}}\right|=0.41\pm0.07\,.
\end{equation}

In the parametrization of the $B^\pm\to \pi^\pm K$ and $B_d\to\pi^\mp K^\pm$
observables, it turns out to be useful to introduce the quantities
\begin{equation}
r\equiv\frac{|T|}{\sqrt{\langle|P|^2\rangle}}\,,\quad\epsilon_{\rm C}\equiv
\frac{|P_{\rm ew}^{\rm C}|}{\sqrt{\langle|P|^2\rangle}}\,,
\end{equation}
with
\begin{equation}
\left\langle|P|^2\right\rangle\equiv\frac{1}{2}\left(|P|^2+
|\overline{P}|^2\right),
\end{equation}
as well as the CP-conserving strong phase differences
\begin{equation}
\delta\equiv\delta_T-\delta_{tc}\,,\quad\Delta_{\rm C}
\equiv\delta_{\rm ew}^{\rm C}-\delta_{tc}\,.
\end{equation}
The CP-conjugate amplitude $\overline{P}$ is obtained from (\ref{Bpampl}) by
simply reversing the sign of the weak phase $\gamma$. A similar comment 
applies also to all other CP-conjugate decay amplitudes appearing in this 
paper. In addition to the ratio $R$ of combined $B\to\pi K$ branching ratios 
defined by  (\ref{Def-R}), also the ``pseudo-asymmetry'' 
\begin{equation}\label{A0-def}
A_0\equiv\frac{\mbox{BR}(B^0_d\to\pi^-K^+)-\mbox{BR}(\overline{B^0_d}\to
\pi^+K^-)}{\mbox{BR}(B^+\to\pi^+K^0)+\mbox{BR}(B^-\to\pi^-\overline{K^0})}=
A_{\rm CP}(B_d\to\pi^\mp K^\pm)\,R
\end{equation}
plays an important role to probe the CKM angle $\gamma$. Explicit expressions
for $R$ and $A_0$ in terms of the parameters specified above are given
in \cite{defan}.

As we already noted, the electroweak penguins are ``colour-suppressed'' 
in the case of the decays $B^+\to\pi^+K^0$ and $B^0_d\to\pi^-K^+$. 
Calculations performed at the perturbative quark level, where
the relevant hadronic matrix elements are treated within the ``factorization''
approach, typically give $\epsilon_{\rm C}={\cal O}(1\%)$ \cite{akl}. 
These crude estimates may, however, underestimate the role of these 
topologies \cite{groro,neubert}. An improved theoretical description of the 
electroweak penguins is possible, using the general expressions for the 
corresponding four-quark operators, appropriate Fierz transformations and
the $SU(2)$ isospin symmetry. Following these lines \cite{defan} 
(see also \cite{PAPIII,rf-ewp3}), we arrive at 
\begin{equation}\label{eps-r-final0}
\left|\frac{P_{\rm ew}^{\rm C}}{T}\right|\,
e^{i\left(\delta_{\rm ew}^{\rm C}-\delta_T\right)}=-\,\frac{3}{2\lambda^2R_b}
\left[\frac{C_9(\mu)+C_{10}(\mu)\zeta(\mu)}{C_1'(\mu)+C_2'(\mu)\zeta(\mu)}
\right],
\end{equation}
with
\begin{equation}\label{zeta-def}
\zeta(\mu)=\frac{\langle K^0\pi^+|Q_2^u(\mu)|B^+\rangle+\langle 
K^+\pi^-|Q_2^u(\mu)|B^0_d\rangle}{\langle K^0\pi^+|Q_1^u(\mu)|B^+\rangle+
\langle K^+\pi^-|Q_1^u(\mu)|B^0_d\rangle}
\end{equation}
and
\begin{equation}
C_1'(\mu)\equiv C_1(\mu)+\frac{3}{2}\,C_9(\mu),\quad C_2'(\mu)\equiv C_2(\mu)+
\frac{3}{2}\,C_{10}(\mu). 
\end{equation}
Here $C_{1,2}(\mu)$ are the Wilson coefficients of the current--current 
operators 
\begin{eqnarray}
Q_1^u&=&(\bar u_{\alpha} s_{\beta})_{{\rm V-A}}
\;(\bar b_{\beta} u_{\alpha})_{{\rm V-A}}\nonumber\\
Q_2^u&=&(\bar u_{\alpha} s_{\alpha})_{{\rm V-A}}\;
(\bar b_{\beta} u_{\beta})_{{\rm V-A}}\,,\label{CC-OP-def}
\end{eqnarray}
and the coefficients $C_{9,10}(\mu)$ are those of the electroweak penguin 
operators 
\begin{eqnarray}
Q_9&=&\frac{3}{2}\,(\bar b_{\alpha}s_{\alpha})_{{\rm V-A}}
\sum\limits_{q=u,d,c,s,b}e_q\,(\bar q_{\beta}q_{\beta})_{{\rm V-A}}\nonumber\\
Q_{10}&=&\frac{3}{2}\,(\bar b_{\alpha}s_{\beta})_{{\rm V-A}}
\sum\limits_{q=u,d,c,s,b}e_q\,
(\bar q_{\beta} q_{\alpha})_{{\rm V-A}}\,.\label{EWP-OP-def}
\end{eqnarray}
As usual, $\alpha$ and $\beta$ are colour indices, and $e_q$ denotes 
the quark charges. It should be kept in mind that two electroweak
penguin operators, $Q_7$ and $Q_8$, with tiny Wilson coefficients, and
electroweak penguins with internal charm- and up-quark exchanges were neglected
in the derivation of (\ref{eps-r-final0}). In our numerical estimates given 
below, it will suffice to use the leading-order values \cite{BBL}
\begin{equation}\label{WC}
C_1(m_b)=-\,0.308,\,\,\,C_2(m_b)=1.144,\,\,\,C_9(m_b)/\alpha=-\,1.280,
\,\,\,C_{10}(m_b)/\alpha=0.328
\end{equation}
with $\alpha=1/129$. It is possible to rewrite (\ref{eps-r-final0}) 
as follows \cite{defan}:
\begin{eqnarray}
\frac{\epsilon_{\rm C}}{r}\,e^{i(\Delta_{\rm C}-\delta)}&=&
\frac{3}{2\lambda^2R_b}\left[\frac{C_1'(\mu)C_9(\mu)-C_2'(\mu)
C_{10}(\mu)}{C_2'^2(\mu)-C_1'^2(\mu)}\right.\nonumber\\
&&\left.+\,a_{\rm C}\,e^{i\omega_{\rm C}}\left\{\frac{C_1'(\mu)C_{10}(\mu)-
C_2'(\mu)C_9(\mu)}{C_2'^2(\mu)-C_1'^2(\mu)}\right\}\right],
\end{eqnarray}
where we will neglect the first, strongly suppressed term
\begin{equation}\label{suppr}
\frac{C_1'(\mu)C_9(\mu)-C_2'(\mu)C_{10}(\mu)}{C_1'(\mu)C_{10}(\mu)-
C_2'(\mu)C_9(\mu)}={\cal O}(10^{-2})
\end{equation}
in the following considerations:
\begin{equation}\label{eps-r-final}
\frac{\epsilon_{\rm C}}{r}\,e^{i(\Delta_{\rm C}-\delta)}\approx
\frac{3}{2\lambda^2R_b}\left[\frac{C_1'(\mu)C_{10}(\mu)-
C_2'(\mu)C_9(\mu)}{C_2'^2(\mu)-C_1'^2(\mu)}\right]
a_{\rm C}\,e^{i\omega_{\rm C}}\,.
\end{equation}
The combination of Wilson coefficients in this expression is essentially 
renormalization-scale-independent and changes only by 
${\cal O}(1\%)$ when evolving from $\mu=M_W$ down to $\mu=m_b$. Employing 
$R_b=0.41$ and the Wilson coefficients given in (\ref{WC}) yields \cite{defan}
\begin{equation}
\frac{\epsilon_{\rm C}}{r}\,e^{i(\Delta_{\rm C}-\delta)}\approx0.66\times 
a_{\rm C}\,e^{i\omega_{\rm C}}.
\end{equation}
The quantity $a_{\rm C}\,e^{i\omega_{\rm C}}$ is given by 
\begin{equation}
a_{\rm C}\,e^{i\omega_{\rm C}}\equiv\frac{a_2^{\rm eff}}{a_1^{\rm eff}}
=\frac{C_1'(\mu)\,\zeta(\mu)+C_2'(\mu)}{C_1'(\mu)+C_2'(\mu)\,\zeta(\mu)}\,,
\end{equation}
where $a_1^{\rm eff}$ and $a_2^{\rm eff}$ correspond to a generalization of
the usual phenomenological ``colour'' factors $a_1$ and $a_2$, describing the 
intrinsic strength of ``colour-suppressed'' and ``colour-allowed'' decay 
processes, respectively \cite{defan}. Note that the ``factorization'' 
approach gives $\zeta(\mu_{\rm F})=3$, where $\mu_{\rm F}$ is the 
``factorization scale''. Comparing experimental data on 
$B^-\to D^{(\ast)0}\pi^-$ and $\overline{B^0_d}\to 
D^{(\ast)+}\pi^-$, as well as on $B^-\to D^{(\ast)0}\rho^-$ and 
$\overline{B^0_d}\to D^{(\ast)+}\rho^-$ decays gives $a_2/a_1={\cal O}(0.25)$. 
Here $a_1$ and $a_2$ are -- in contrast to $a_1^{\rm eff}$ and 
$a_2^{\rm eff}$ -- real quantities, and their relative sign is found to 
be positive. Experimental studies of $B\to J/\psi\, K^{(\ast)}$ decays 
favour also $|a_2/a_1|={\cal O}(0.25)$. If we assume that the strength of 
``colour suppression'' in $B\to\pi K$ decays is of the same order of
magnitude, i.e.\ $a_{\rm C}=0.25$, we obtain a value of $\epsilon_{\rm C}/r$ 
that is larger by a factor of 3 than the ``factorized'' result
\begin{equation}\label{factor}
\left.\frac{\epsilon_{\rm C}}{r}\,e^{i(\Delta_{\rm C}-\delta)}
\right|_{\rm fact}=\,0.06\times\left[\frac{0.41}{R_b}\right],
\end{equation}
corresponding to $\mu=\mu_{\rm F}$ and $\zeta(\mu_{\rm F})=3$ in 
(\ref{eps-r-final0}). The expression (\ref{eps-r-final}) shows nicely that
the usual terminology of ``colour-suppressed'' electroweak penguins in 
(\ref{ampl-rel1}) is justified, since $P_{\rm ew}^{\rm C}$ is proportional
to the generalized ``colour'' factor $a_2^{\rm eff}$. Moreover, it implies
a correlation between $\epsilon_{\rm C}$ and $r$, which is given by 
\begin{equation}\label{eps-r-rel}
\epsilon_{\rm C}=q_{\rm C}\,r,\quad \Delta_{\rm C}=\delta+\omega_{\rm C}
\end{equation}
with 
\begin{equation}
q_{\rm C}\approx0.66\times\left[\frac{0.41}{R_b}\right]\times a_{\rm C}. 
\end{equation}

The ratio $R$ defined by (\ref{Def-R}) can be expressed as 
follows \cite{defan}:
\begin{equation}\label{R-exp}
R=1-\frac{2\,r}{u}\left(h\cos\delta+k\sin\delta\right)+v^2r^2,
\end{equation}
where
\begin{eqnarray}
h&=&\cos\gamma+\rho\cos\theta-q_{\rm C}\left[\,\cos\omega_{\rm C}+
\rho\cos(\theta-\omega_{\rm C})\cos\gamma\,\right]\label{h-def}\\
k&=&\rho\sin\theta+q_{\rm C}\left[\,\sin\omega_{\rm C}-\rho\sin(\theta-
\omega_{\rm C})\cos\gamma\,\right]
\end{eqnarray}
and
\begin{eqnarray}
u&=&\sqrt{1+2\,\rho\,\cos\theta\cos\gamma+\rho^2}\\
v&=&\sqrt{1-2\,q_{\rm C}\cos\omega_{\rm C}\cos\gamma+
q_{\rm C}^2}\,.\label{v-def}
\end{eqnarray}
The pseudo-asymmetry $A_0$ (see (\ref{A0-def})) takes the form
\begin{equation}\label{A0-exp}
A_0=A_++2\,\frac{r}{u}\,\left[\,\sin\delta+q_{\rm C}\,\rho\,
\sin(\delta-\theta+\omega_{\rm C})\,\right]\,\sin\gamma-2\,q_{\rm C}\,r^2\,
\sin\omega_{\rm C}\,\sin\gamma,
\end{equation}
where 
\begin{equation}\label{Ap-def}
A_+\equiv\frac{\mbox{BR}(B^+\to\pi^+K^0)-\mbox{BR}(B^-\to\pi^-
\overline{K^0})}{\mbox{BR}(B^+\to\pi^+K^0)+\mbox{BR}(B^-\to\pi^-
\overline{K^0})}=-\,\frac{2\,\rho\,\sin\theta\sin\gamma}{1+
2\,\rho\,\cos\theta\cos\gamma+\rho^2}
\end{equation}
measures direct CP violation in the decay $B^+\to\pi^+K^0$. Note that tiny 
phase-space effects have been neglected in (\ref{R-exp}) and (\ref{A0-exp})
(for a more detailed discussion, see  \cite{fm2}). 

\boldmath
\subsection{The Decays $B^\pm\to\pi^0 K^\pm$ and 
$B_d\to\pi^0 K$}\label{SubSec22}
\unboldmath
Let us now turn to the decays $B^+\to\pi^0K^+$, $B^0_d\to\pi^0K^0$ and their
charge conjugates. The $SU(2)$ isospin symmetry implies the following 
amplitude relation \cite{nq,ghlr}:
\begin{eqnarray}
A(B^+\to\pi^+K^0)\,+\,\sqrt{2}\,A(B^+\to\pi^0K^+)&=&
\sqrt{2}\,A(B^0_d\to\pi^0K^0)\,+\,A(B^0_d\to\pi^-K^+)\nonumber\\
=-\left[(T+C)\,+\,P_{\rm ew}\right]&\equiv& 3\, A_{3/2},\label{ampl-rel2}
\end{eqnarray}
where $P_{\rm ew}$ is due to electroweak penguins and $A_{3/2}$ 
refers to a $\pi K$ isospin configuration with $I=3/2$. Note that there is 
no $I=1/2$ component present in  (\ref{ampl-rel2}). Since we have
\begin{equation}\label{TC-def}
T+C=|T+C|\,e^{i\delta_{T+C}}\,e^{i\gamma}
\end{equation}
and
\begin{equation}\label{Pew-def}
P_{\rm ew}=-\,|P_{\rm ew}|e^{i\delta_{\rm ew}},
\end{equation}
the phase structure of the amplitude relation (\ref{ampl-rel2}) is completely 
analogous to the one given in (\ref{ampl-rel1}). We just have to perform
the replacements 
\begin{equation}
T\to T+C\quad \mbox{and}\quad P_{\rm ew}^{\rm C}\to P_{\rm ew}.
\end{equation}
The notation of $T+C$ reminds us that this amplitude receives 
contributions both from ``colour-allowed'' and from ``colour-suppressed'' 
$\bar b\to\bar uu\bar s$ tree-diagram-like topologies \cite{grl}. A
similar comment applies to the electroweak penguin amplitude $P_{\rm ew}$,
receiving also contributions both from ``colour-allowed'' and from 
``colour-suppressed'' electroweak penguin topologies \cite{ghlr}. 
If we neglect electroweak penguin topologies with internal
charm and up quarks, as well as the electroweak penguin operators $Q_7$ and 
$Q_8$, which have tiny Wilson coefficients, perform appropriate Fierz 
transformations of the remaining electroweak penguin operators $Q_9$ and 
$Q_{10}$ and, moreover, apply the $SU(2)$ isospin symmetry, we arrive at
\begin{equation}\label{PEW1}
\left|\frac{P_{\rm ew}}{T+C}\right|\,
e^{i\left(\delta_{\rm ew}-\delta_{T+C}\right)}=-\,\frac{3}{2\lambda^2R_b}
\left[\frac{C_9(\mu)+C_{10}(\mu)\tilde\zeta(\mu)}{C_1'(\mu)+
C_2'(\mu)\tilde\zeta(\mu)}\right],
\end{equation}
with
\begin{eqnarray}
\tilde\zeta(\mu)&=&\frac{\langle K^0\pi^+|Q_2^u(\mu)|B^+\rangle+
\sqrt{2}\,\langle K^+\pi^0|Q_2^u(\mu)|B^+\rangle}{\langle K^0\pi^+|
Q_1^u(\mu)|B^+\rangle+\sqrt{2}\,\langle K^+\pi^0|Q_1^u(\mu)|B^+
\rangle}\nonumber\\
&=&\frac{\sqrt{2}\,\langle K^0\pi^0|Q_2^u(\mu)|B^0_d\rangle+
\langle K^+\pi^-|Q_2^u(\mu)|B^0_d\rangle}{\sqrt{2}\,\langle K^0\pi^0|
Q_1^u(\mu)|B^0_d\rangle+\langle K^+\pi^-|Q_1^u(\mu)|B^0_d\rangle}
\equiv\frac{\left\langle Q_2^u(\mu)\right\rangle}{\left\langle Q_1^u(\mu)
\right\rangle},
\end{eqnarray}
which is completely analogous to (\ref{eps-r-final0}) and (\ref{zeta-def}). 
Since the $SU(3)$ flavour symmetry of strong interactions implies 
\begin{equation}
\left\langle Q_1^u(\mu)\right\rangle=\left\langle Q_2^u(\mu)\right\rangle,
\end{equation}
it is useful to rewrite (\ref{PEW1}) as follows:
\begin{equation}\label{EE1}
\left|\frac{P_{\rm ew}}{T+C}\right|\,e^{i(\delta_{\rm ew}-\delta_{T+C})}=
-\,\frac{3}{2\lambda^2R_b}\left[\frac{C_9(\mu)+C_{10}(\mu)+
\left\{C_9(\mu)-C_{10}(\mu)\right\}\zeta_{SU(3)}(\mu)}{C_1'(\mu)+C_2'(\mu)+
\left\{C_1'(\mu)-C_2'(\mu)\right\}\zeta_{SU(3)}(\mu)}\right],
\end{equation}
where
\begin{eqnarray}
\zeta_{SU(3)}(\mu)&=&\frac{1-\tilde\zeta(\mu)}{1+\tilde\zeta(\mu)}=
\frac{\langle K^0\pi^+|[Q_1^u(\mu)-Q_2^u(\mu)]
|B^+\rangle+\sqrt{2}\,\langle K^+\pi^0|[Q_1^u(\mu)-Q_2^u(\mu)]|
B^+\rangle}{\langle K^0\pi^+|[Q_1^u(\mu)+Q_2^u(\mu)]|B^+\rangle+\sqrt{2}\,
\langle K^+\pi^0|[Q_1^u(\mu)+Q_2^u(\mu)]|B^+\rangle}\nonumber\\
&&=\frac{\sqrt{2}\,\langle K^0\pi^0|[Q_1^u(\mu)-Q_2^u(\mu)]
|B^0_d\rangle+\langle K^+\pi^-|[Q_1^u(\mu)-Q_2^u(\mu)]|
B^0_d\rangle}{\sqrt{2}\,\langle K^0\pi^0|[Q_1^u(\mu)+Q_2^u(\mu)]|B^0_d\rangle+
\langle K^+\pi^-|[Q_1^u(\mu)+Q_2^u(\mu)]|B^0_d\rangle}
\end{eqnarray}
describes $SU(3)$-breaking corrections. In the strict $SU(3)$ limit, we
have $\zeta_{SU(3)}(\mu)=0$, and obtain \cite{nr1}
\begin{eqnarray}
\lefteqn{\left|\frac{P_{\rm ew}}{T+C}\right|\,
e^{i(\delta_{\rm ew}-\delta_{T+C})}\equiv q\,e^{i\omega}
\approx-\,\frac{3}{2\lambda^2R_b}\left[\frac{C_9(\mu)+
C_{10}(\mu)}{C_1'(\mu)+C_2'(\mu)}\right]}\nonumber\\
&&\quad\approx\frac{3}{2\lambda^2R_b}\left[\frac{C_1'(\mu)C_{10}(\mu)-
C_2'(\mu)C_9(\mu)}{C_2'^2(\mu)-C_1'^2(\mu)}\right]=0.66\times
\left[\frac{0.41}{R_b}\right],\label{EE3}
\end{eqnarray}
which is related to $q_{\rm C}\,e^{i\omega_{\rm C}}$ through 
\begin{equation}
q_{\rm C}\,e^{i\omega_{\rm C}}\approx q\,e^{i\omega}\times a_{\rm C}\,.
\end{equation}
Here we have again neglected the strongly suppressed term (\ref{suppr}).
Within the ``factorization'' approximation, we have very small 
$SU(3)$-breaking corrections at the level of a few per cent \cite{nr1} 
(see also  \cite{PAPIII}). Unfortunately, we have no insights into 
non-factorizable $SU(3)$ breaking at present. Taking into account both the
factorizable corrections, which shift $q$ from 0.66 to 0.63, and 
the present experimental uncertainty of $R_b$ (see (\ref{CKM-exp})), 
Neubert and Rosner give the range of $q=0.63\pm0.15$ \cite{nr1}.

If we compare (\ref{ampl-rel2}) with (\ref{ampl-rel1}), we find that
the observables of the charged $B$-meson decays $B^\pm\to\pi^\pm K$,
$\pi^0K^\pm$ corresponding to $R$ and $A_0$ have to be defined as follows:
\begin{eqnarray}
R_{\rm c}&\equiv&2\left[\frac{\mbox{BR}(B^+\to\pi^0K^+)+
\mbox{BR}(B^-\to\pi^0K^-)}{\mbox{BR}(B^+\to\pi^+K^0)+
\mbox{BR}(B^-\to\pi^-\overline{K^0})}\right]\label{Rc-def}\\
A_0^{\rm c}&\equiv&2\left[\frac{\mbox{BR}(B^+\to\pi^0K^+)-
\mbox{BR}(B^-\to\pi^0K^-)}{\mbox{BR}(B^+\to\pi^+K^0)+
\mbox{BR}(B^-\to\pi^-\overline{K^0})}\right]=A_{\rm CP}(B^\pm\to\pi^0K^\pm)\,
R_{\rm c}\,.\label{A0c-def}
\end{eqnarray}
Concerning strategies to probe the CKM angle $\gamma$, the ratio 
$R_{\rm c}$ is more convenient than the quantity $R_*=1/R_{\rm c}$, which
was considered by Neubert and Rosner in \cite{nr1}. The preliminary 
results on the CP-averaged branching ratios 
\begin{eqnarray}
\mbox{BR}(B^\pm\to\pi^0K^\pm)&=&(1.5\pm0.4\pm0.3)\times10^{-5}\\
\mbox{BR}(B^\pm\to\pi^\pm K)&=&(1.4\pm0.5\pm0.2)\times10^{-5},
\end{eqnarray}
which, very recently, were reported by the CLEO collaboration \cite{newCLEO}, 
give
\begin{equation}\label{Rc-range}
R_{\rm c}=2.1\pm1.1\,.
\end{equation}
Here we have added the errors in quadrature. This result differs 
significantly from the present value of $R=1.0\pm0.4$, although the 
uncertainties are too large to say anything definite. 

In the case of the neutral modes $B_d\to\pi^0 K$, $\pi^\mp K^\pm$, 
we have
\begin{eqnarray}
R_{\rm n}&\equiv&\frac{1}{2}\left[\frac{\mbox{BR}(B^0_d\to\pi^-K^+)+
\mbox{BR}(\overline{B^0_d}\to\pi^+K^-)}{\mbox{BR}(B^0_d\to\pi^0K^0)+
\mbox{BR}(\overline{B^0_d}\to\pi^0\overline{K^0})}\right]\label{Rn-def}\\
A_0^{\rm n}&\equiv&\frac{1}{2}\left[\frac{\mbox{BR}(B^0_d\to\pi^-K^+)-
\mbox{BR}(\overline{B^0_d}\to\pi^+K^-)}{\mbox{BR}(B^0_d\to\pi^0K^0)+
\mbox{BR}(\overline{B^0_d}\to\pi^0\overline{K^0})}\right]=A_{\rm CP}
(B_d\to\pi^\mp K^\pm)\,R_{\rm n}.\label{A0n-def}
\end{eqnarray}
While the CLEO collaboration recently reported the preliminary 
result \cite{newCLEO}
\begin{equation}
\mbox{BR}(B_d\to\pi^\mp K^\pm)=(1.4\pm0.3\pm0.2)\times10^{-5},
\end{equation}
there is at present only an upper limit available for the decay 
$B_d\to\pi^0K$, which is given by BR$(B_d\to\pi^0K)<4.1\times10^{-5}$ 
\cite{cleo}.

The parametrization of the observables $R_{\rm c}$, $A_0^{\rm c}$ and 
$R_{\rm n}$, $A_0^{\rm n}$ is completely analogous to  (\ref{R-exp})
and (\ref{A0-exp}) and can be obtained straightforwardly from these
expressions by performing appropriate replacements. The most obvious one
is the following:
\begin{equation}\label{q-repl}
q_{\rm C}\,e^{i\omega_{\rm C}}\to q\,e^{i\omega}.
\end{equation}
Moreover, we have to substitute
\begin{equation}
r\to r_{\rm c}\equiv\frac{|T+C|}{\sqrt{\langle|P|^2\rangle}}\,, \quad
\delta\to \delta_{\rm c}\equiv\delta_{T+C}-\delta_{tc}
\end{equation}
in the case of the observables $R_{\rm c}$ and $A_0^{\rm c}$. The parameter 
$\rho\,e^{i\theta}$, which is defined through the $B^+\to\pi^+K^0$
decay amplitude, remains unchanged. This is in contrast to the case of
the neutral modes $B_d\to\pi^0 K$, $\pi^\mp K^\pm$. Here the decay 
$B_d^0\to\pi^0 K^0$ takes the role of the mode $B^+\to\pi^+K^0$. In 
analogy to  (\ref{Bpampl}), its decay amplitude can be expressed as
\begin{equation}\label{Bnampl}
\sqrt{2}\,A(B^0_d\to\pi^0K^0)\equiv P_{\rm n}=
-\left(1-\frac{\lambda^2}{2}\right)\lambda^2A
\left[1+\rho_{\rm n}\,e^{i\theta_{\rm n}}e^{i\gamma}\right]
{\cal P}_{tc}^{\rm n}\,,
\end{equation}
where $\rho_{\rm n}\,e^{i\theta_{\rm n}}$ takes the form 
\begin{equation}\label{rho-n-def}
\rho_{\rm n}\,e^{i\theta_{\rm n}}=\frac{\lambda^2R_b}{1-\lambda^2/2}
\left[1-\left(\frac{{\cal P}_{uc}^{\rm n}-{\cal C}}{{\cal P}_{tc}^{\rm n}}
\right)\right].
\end{equation}
Here ${\cal P}_{tc}^{\rm n}\equiv|{\cal P}_{tc}^{\rm n}|\,
e^{i\delta_{tc}^{\rm n}}$ and ${\cal P}_{uc}^{\rm n}$ correspond to 
differences of penguin topologies with internal top and charm and up and 
charm quarks, respectively (see (\ref{Ptc})). In contrast to the 
$B^+\to\pi^+K^0$ case, these quantities receive contributions also from 
``colour-allowed'' electroweak penguin topologies. The amplitude ${\cal C}$ 
is due to insertions of the current--current operators (\ref{CC-OP-def}) 
into ``colour-suppressed'' tree-diagram-like topologies. In order to 
parametrize the observables $R_{\rm n}$ and $A_0^{\rm n}$ with the help of  
(\ref{R-exp}) and (\ref{A0-exp}), we have -- in addition to 
(\ref{q-repl}) -- to perform the following replacements:
\begin{equation}
r\to r_{\rm n}\equiv\frac{|T+C|}{\sqrt{\langle|P_{\rm n}|^2\rangle}}\,, 
\quad\delta\to \delta_{\rm n}\equiv\delta_{T+C}-\delta_{tc}^{\rm n}\,,\quad
\rho\,e^{i\theta}\to\rho_{\rm n}\,e^{i\theta_{\rm n}}\,.
\end{equation}

\newpage

\boldmath
\section{Probing the CKM Angle $\gamma$ with the Decays\\ 
$B^\pm\to\pi^\pm K$ and $B_d\to\pi^\mp K^\pm$}\label{PAPIII-strat}
\unboldmath 
Before we turn to strategies to constrain and determine the CKM angle
$\gamma$ with the help of the charged decays $B^\pm\to\pi^\pm K$, 
$\pi^0K^\pm$ in Section~\ref{char-strat}, and to a new approach dealing
with the neutral modes $B_d\to\pi^0K$, $\pi^\mp K^\pm$ in 
Section~\ref{neut-strat}, let us recapitulate in this section the methods 
using the decays $B^\pm\to\pi^\pm K$ and $B_d\to\pi^\mp K^\pm$. This will
allow us, later, to better compare the virtues and weaknesses of all
three approaches. Moreover, it is useful to reanalyse the $B^\pm\to\pi^\pm K$ 
and $B_d\to\pi^\mp K^\pm$ modes in the light of the most recent CLEO results
\cite{newCLEO}, thereby pointing out some interesting features that had not
been emphasized in previous work \cite{ICHEP98}. 

\boldmath
\subsection{Strategies to Constrain the CKM Angle $\gamma$}
\unboldmath
Before turning to strategies to extract $\gamma$, let us first focus on 
methods to constrain this angle through the ratio $R$ of the combined 
$B_d\to\pi^\mp K^\pm$ and $B^\pm\to\pi^\pm K$ branching ratios introduced in
(\ref{Def-R}), i.e.\ without making use of the expected sizeable CP 
asymmetry arising in $B_d\to\pi^\mp K^\pm$. At present, CP-violating 
effects in $B\to\pi K$ decays have not yet been observed, and only data 
for the corresponding combined, i.e.\ CP-averaged, branching ratios are 
available.

In order to constrain the CKM angle $\gamma$ with the help of the observable
$R$, we keep the CP-conserving strong phase $\delta$, which is under no 
theoretical control and completely unknown at present, as a free parameter 
\cite{fm2}. Using the general expression (\ref{R-exp}), we find that $R$ 
takes the following extremal values:
\begin{equation}\label{R-bounds0}
\left.R_{\rm min}^{\rm max}\right|_{\delta}=1\,\pm\,2\,\frac{r}{u}\,
\sqrt{h^2+k^2}\,+\,v^2r^2\,,
\end{equation}
which constrain $\gamma$, provided $r$ can be determined (in \cite{fm2}, a 
different approach was used to derive these constraints for the special
case of neglected rescattering and electroweak penguin effects, i.e.\ for
$\rho=q_{\rm C}=0$). In the case of the decays $B^\pm\to\pi^\pm K$ and 
$B_d\to\pi^\mp K^\pm$, flavour symmetry arguments are not sufficient to 
fix the parameter $r$ -- this is in contrast to the case of $r_{\rm c,\,n}$ 
of the ``charged'' and ``neutral'' strategies discussed in the following 
sections -- and an additional input, for example ``factorization'' or the 
neglect of ``colour-suppressed'' topologies in the decay $B^+\to\pi^+\pi^0$, 
have to be used to accomplish this task. Following these lines, present data 
give $r=0.15\pm0.05$ \cite{groro-rel}. Since the properly defined amplitude 
$T$, which governs the parameter $r$, is not just a ``tree'' amplitude, but 
receives contributions also from certain penguin and annihilation topologies 
\cite{defan,bfm}, it is at present difficult to estimate the theoretical 
uncertainty of $r$ in a realistic way. Optimistic analyses come to the 
conclusion that a future theoretical uncertainty of $\Delta r={\cal O}(10\%)$ 
may be achievable \cite{groro,wuegai}. However, if rescattering processes 
of the kind $B^+\to\{\pi^0K^+\}\to\pi^+K^0$ should play an important role, 
the uncertainties may be significantly larger. Consequently, it would be 
favourable to have constraints on $\gamma$ that do not depend on $r$. 

\begin{figure}
\centerline{
\rotate[r]{
\epsfxsize=9.2truecm
\epsffile{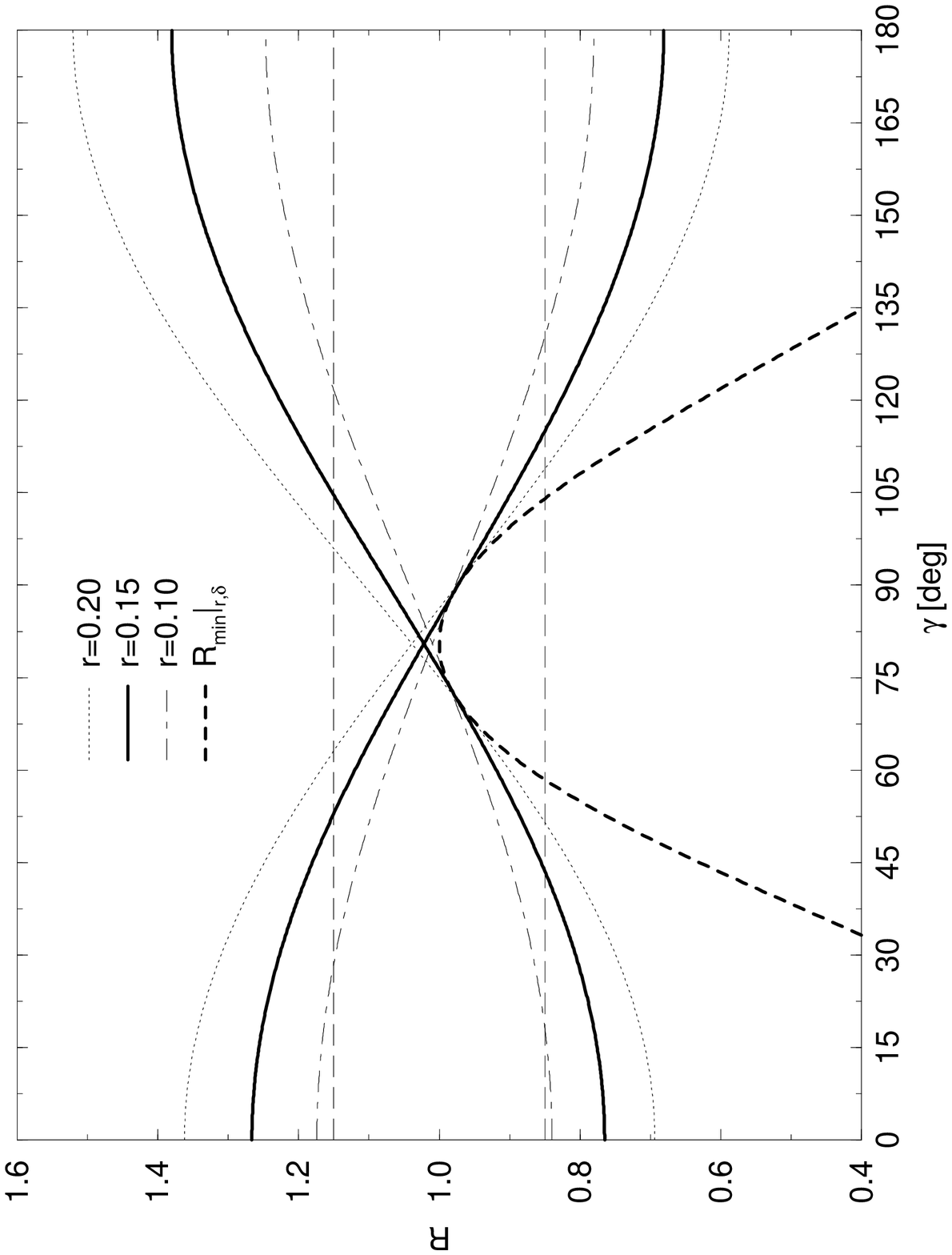}}}
\caption{The dependence of the extremal values of $R$ given in 
(\ref{R-bounds0}) and (\ref{RminEWP0}) on the CKM angle $\gamma$ for 
$q_{\rm C}\,e^{i\omega_{\rm C}}=0.66\times0.25$ in the case of negligible 
rescattering effects, i.e.\ $\rho=0$.}\label{fig:FM-bounds}
\end{figure}

\begin{figure}
\centerline{
\rotate[r]{
\epsfxsize=9.2truecm
\epsffile{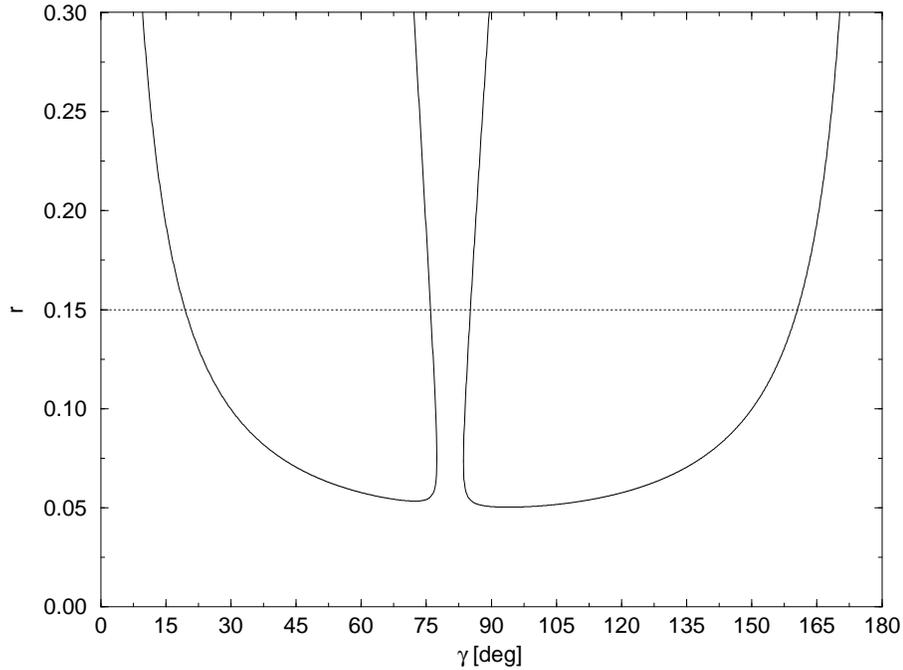}}}
\caption{The contours in the $\gamma$--$r$ plane corresponding to 
$R=1.00$, $A_0=9.96\%$ and $q_{\rm C}\,e^{i\omega_{\rm C}}=0.66\times0.25$
in the case of negligible rescattering effects, i.e.\ 
$\rho=0$.}\label{fig:FM-contours}
\end{figure}

It was pointed out in \cite{fm2} that such bounds can be obtained, provided 
$R$ is found to be smaller than 1. Within our formalism, they can be derived 
by keeping both $\delta$ and $r$ as free parameters in the general
expression (\ref{R-exp}) for $R$. Following these lines, we find that $R$ 
takes the minimal value \cite{defan}
\begin{equation}\label{RminEWP0}
\left.R_{\rm min}\right|_{r,\,\delta}=\left[\frac{1+2\,q_{\rm C}\,\rho\,
\cos(\theta+\omega_{\rm C})+q_{\rm C}^2\,\rho^2}{\left(1-
2\,q_{\rm C}\,\cos\omega_{\rm C}\cos\gamma+q_{\rm C}^2\right)
\left(1+2\,\rho\,\cos\theta\cos\gamma+\rho^2\right)}\right]\sin^2\gamma\,,
\end{equation}
which corresponds to a generalization of the result derived in \cite{fm2} 
(see (\ref{gamma-bound1}) and (\ref{gam0})), and would exclude a certain range 
of $\gamma$ around $90^\circ$, if $R$ is found to be smaller than 1. 
This feature led to great excitement in the $B$-physics community, since
the first results reported by the CLEO collaboration gave $R=0.65\pm0.40$ 
\cite{cleo}. Unfortunately, a recent, preliminary update yields 
$R=1.0\pm0.4$ and is therefore not as promising \cite{newCLEO}, although 
it is too early to draw definite conclusions. 

In Fig.\ \ref{fig:FM-bounds}, we have chosen $q_{\rm C}\,e^{i\omega_{\rm C}}=
0.66\times0.25$ and $\rho=0$ in order to illustrate the dependence of 
(\ref{R-bounds0}) and (\ref{RminEWP0}) on the CKM angle $\gamma$. For 
$R=0.85$, the latter expression would exclude the range of 
$58^\circ\leq\gamma\leq104^\circ$. The values of $r$ used to evaluate 
(\ref{R-bounds0}) correspond to the presently allowed range given by Gronau 
and Rosner \cite{groro-rel}. In the future, the corresponding uncertainty 
of $33\%$ may be reduced by a factor of ${\cal O}(3)$, provided rescattering 
processes play a negligible role. On the other hand, $r$ may in principle be 
shifted significantly, if rescattering effects should turn out to be large. 
Important indicators for this unfortunate case would be large direct CP 
violation in $B^\pm\to\pi^\pm K$ modes, and the size of the branching ratios 
of the decays $B^\pm\to K^\pm K$ and $B_d\to K^+K^-$ 
\cite{defan,rf-FSI,groro-FSI}. 
In order to illustrate the constraints on $\gamma$ in more detail, let us 
assume that $B^\pm\to K^\pm K$ and $B_d\to K^+K^-$ indicate that rescattering 
effects play a very minor role and that the strategies to fix $r$ (see, for 
example, \cite{groro,groro-rel,wuegai}) yield $r=0.15$. As can be read 
off from Fig.~\ref{fig:FM-bounds}, the minimal value of $R$ given in 
(\ref{R-bounds0}) would exclude the range of 
$44^\circ\leq\gamma\leq115^\circ$ in the case of 
$R=0.85$. If we assume that $R$ is found to be equal to 1.15, (\ref{RminEWP0}) 
would not be effective. However, the maximal value of $R$ given in 
(\ref{R-bounds0}) would exclude the range of 
$53^\circ\leq\gamma\leq105^\circ$.

\boldmath
\subsection{Strategies to Determine the CKM Angle $\gamma$}
\unboldmath
As soon as CP violation in $B_d\to\pi^\mp K^\pm$ decays is observed, it is
possible to go beyond the bounds on $\gamma$ discussed in the previous 
subsection. Then we are in a position to eliminate the strong phase
$\delta$ in $R$ with the help of the pseudo-asymmetry $A_0$, thereby fixing 
contours in the $\gamma$--$r$ plane, which are a mathematical implementation
of the simple triangle construction proposed in \cite{PAPIII}. The 
corresponding formulae are quite complicated and are given in \cite{defan}. 
In order to illustrate these contours in a quantitative way, let us 
assume -- in analogy to an example discussed by Neubert and Rosner 
in \cite{nr2} -- that $\gamma=76^\circ$, $r=0.15$ and $\delta=20^\circ$. 
If we use, moreover, $q_{\rm C}\,e^{i\omega_{\rm C}}=0.66\times0.25$ and 
$\rho=0$, we obtain $R=1.00$ and $A_0=9.96\%$. The contours in the 
$\gamma$--$r$ plane corresponding to these ``measured'' observables are 
shown in Fig.\ \ref{fig:FM-contours}. For $r=0.15$, which is represented in
this figure by the dotted line, we have four solutions for $\gamma$: 
$19^\circ$, $76^\circ$, which is the ``true'' value in our example, 
$85^\circ$ and $161^\circ$. Moreover, a range of $78^\circ\leq\gamma\leq
84^\circ$ is excluded. Since the values of $19^\circ$ and $161^\circ$
are outside the presently allowed range of 
$41^\circ\mathrel{\hbox{\rlap{\hbox{\lower4pt\hbox{$\sim$}}}\hbox{$<$}}}
\gamma\mathrel{\hbox{\rlap{\hbox{\lower4pt\hbox{$\sim$}}}\hbox{$<$}}}
97^\circ$ \cite{UTfits}, which is implied by the usual fits of the unitarity 
triangle, we are left with the two ``physical'' solutions of $76^\circ$ and 
$85^\circ$. In this example, the contours in the $\gamma$--$r$ plane have 
the very interesting feature that these solutions are almost independent of 
the value of $r$ (see also~\cite{defan}). Consequently, they are only 
affected to a small extent by the uncertainty of $r$. As we have already 
noted, if rescattering processes should play an important role, it may be 
difficult to fix this parameter in a reliable way. While it is possible to 
take into account the shift of the contours in the $\gamma$--$r$ plane 
due to large rescattering effects, with the help of the decays 
$B^\pm\to K^\pm K$ and the $SU(3)$ flavour symmetry \cite{defan,rf-FSI}, 
there is unfortunately no straightforward approach to accomplish this 
task also in the determination of $r$. In \cite{defan}, also the 
uncertainties related to the ``colour-suppressed'' electroweak penguins
were analysed. If we used, for example, the strongly suppressed 
``factorized'' result (\ref{factor}) to deal with these topologies in the 
contours in the $\gamma$--$r$ plane, we would obtain the solutions 
$\gamma=19^\circ$, $82^\circ$, $91^\circ$, $161^\circ$ for $r=0.15$, i.e.\
our ``physical'' solutions from above would be shifted by $6^\circ$ towards
larger values of $\gamma$. 

This example shows nicely that the new central value of $R=1$ reported 
recently by the CLEO collaboration~\cite{newCLEO} does by no means 
imply -- even if confirmed by future data -- that the modes 
$B^\pm\to\pi^\pm K$ and $B_d\to\pi^\mp K^\pm$ are ``useless'' to probe 
the CKM angle $\gamma$. Although the constraints on this angle that are 
implied by the combined branching ratios of these modes would not be 
effective in this case (see Fig.\ \ref{fig:FM-bounds}), the prospects to 
determine $\gamma$ as soon as CP violation in $B_d\to\pi^\mp K^\pm$ 
is measured appear to be promising.

\boldmath
\section{Probing the CKM Angle $\gamma$ with the Charged\\
Decays $B^\pm\to\pi^\pm K$ and $B^\pm\to\pi^0 K^\pm$}\label{char-strat}
\unboldmath 
The subjects of this section are strategies to probe the CKM angle $\gamma$
with the help of the observables $R_{\rm c}$ and $A_0^{\rm c}$ defined
in (\ref{Rc-def}) and (\ref{A0c-def}). In this context, an important 
additional ingredient is provided by the fact that the amplitude $T+C$ 
can be determined with the help of the decay $B^+\to\pi^+\pi^0$ by 
using the $SU(3)$ flavour symmetry of strong interactions~\cite{grl}:
\begin{equation}\label{T-C-det}
T+C\approx-\,\sqrt{2}\,\frac{V_{us}}{V_{ud}}\,
\frac{f_K}{f_{\pi}}\,A(B^+\to\pi^+\pi^0).
\end{equation}
Here the ratio $f_K/f_{\pi}=1.2$ of the kaon and pion decay constants
takes into account factorizable $SU(3)$-breaking corrections. At present,
the non-factorizable corrections to (\ref{T-C-det}) cannot be treated
in a quantitative way. It should be noted that electroweak penguin
contributions are also not included in this expression. However, the
formalism discussed in Subsection~\ref{SubSec22} applies also to the 
$B\to\pi\pi$ case, where the $SU(2)$ isospin symmetry suffices to derive 
the following expression:
\begin{equation}\label{bdEWP}
\left[\left|\frac{P_{\rm ew}}{T+C}\right|e^{i(\delta_{\rm ew}-
\delta_{T+C})}\right]_{\bar b\to\bar d}=\frac{3}{2 R_b}\left[\frac{C_9(\mu)+
C_{10}(\mu)}{C_1'(\mu)+C_2'(\mu)}\right]=-\,3.3\times
\left[\frac{0.41}{R_b}\right]\times10^{-2}.
\end{equation}
As in the $\bar b\to\bar s$ case, the amplitues $(T+C)_{\bar b\to\bar d}$
and $(P_{\rm ew})_{\bar b\to\bar d}$ are proportional to the CKM factors
$\lambda_u^{(d)}$ and $\lambda_c^{(d)}$, respectively. Using (\ref{bdEWP}),
we find the corrected expression
\begin{equation}\label{T-C-EWdet}
T+C\approx-\,\sqrt{2}\,\frac{V_{us}}{V_{ud}}\,
\frac{f_K}{f_{\pi}}\,\left[\frac{A(B^+\to\pi^+\pi^0)}{1+0.033\times 
e^{-i\gamma}}\right].
\end{equation}
Consequently, the electroweak penguins lead to a correction to  
(\ref{T-C-det}) that is at most a few per cent. It is interesting to note
that a theoretical input similar to (\ref{bdEWP}) allows us to include
electroweak penguin topologies in the well-known Gronau--London
method \cite{gl} to determine the angle $\alpha$ of the unitarity triangle 
with the help of $B\to\pi\pi$ isospin relations. This by-product of our
considerations is discussed in more detail in the appendix.

\boldmath
\subsection{Strategies to Constrain the CKM Angle $\gamma$}
\unboldmath
The constraints on $\gamma$ implied by (\ref{R-bounds0}) and (\ref{RminEWP0})
apply also to the $B^\pm\to\pi^\pm K$, $\pi^0 K^\pm$ case, if straightforward
replacements are performed. We just have to substitute
\begin{equation}
R\to R_{\rm c}\,,\quad r\to r_{\rm c}\,,\quad 
q_{\rm C}\,e^{i\omega_{\rm C}}\to q\,e^{i\omega}
\end{equation}
in these expressions, leading to the extremal values
\begin{equation}\label{Rc-bounds}
\left.R_{\rm c}^{\rm ext}
\right|_{\delta_{\rm c}}=1\,\pm\,2\,\frac{r_{\rm c}}{u}\,
\sqrt{h^2+k^2}\,+\,v^2r^2_{\rm c}
\end{equation}
and
\begin{equation}\label{RcminEWP}
\left.R_{\rm c}^{\rm min}\right|_{r_{\rm c},\delta_{\rm c}}=
\left[\frac{1+2\,q\,\rho\,\cos(\theta+\omega)+q^2\rho^2}{\left(1-
2\,q\,\cos\omega\cos\gamma+q^2\right)
\left(1+2\,\rho\,\cos\theta\cos\gamma+\rho^2\right)}\right]\sin^2\gamma\,.
\end{equation}
Note that also $q_{\rm C}\,e^{i\omega_{\rm C}}$ has to be replaced by 
$q\,e^{i\omega}$ in the quantities $h$, $k$ and $v$ specified in 
(\ref{h-def})--(\ref{v-def}).  In comparison with $B^\pm\to\pi^\pm K$ and 
$B_d\to\pi^\mp K^\pm$, the decays $B^\pm\to\pi^\pm K$, $\pi^0 K^\pm$ 
have the advantage that $r_{\rm c}$ can be extracted with the help of 
(\ref{T-C-det}), i.e.\ by using only the $SU(3)$ flavour symmetry. 
The present data give $r_{\rm c}=0.24\pm0.06$ \cite{nr1}. 

Because of the present experimental range of $R_{\rm c}^{\rm exp}=2.1\pm1.1$, 
the bounds on $\gamma$ associated with (\ref{RcminEWP}) are not effective at
the moment and the major role to constrain this angle is played by the maximal 
value of $R_{\rm c}$, which corresponds to ``+'' in (\ref{Rc-bounds}). The 
values of $\gamma$ implying $R_{\rm c}^{\rm exp}>R_{\rm c}^{\rm max}$ are 
excluded. These constraints correspond to the bound pointed out by Neubert 
and Rosner in \cite{nr1}, who considered the observable $R_*=1/R_{\rm c}$ 
and performed an expansion in the parameter $r_{\rm c}$ in order to derive 
their bound. Moreover, rescattering effects were completely neglected, i.e.\ 
only the case $\rho=0$ was considered, and $\omega=0^\circ$ was assumed. 
In contrast, our result (\ref{Rc-bounds}) is valid {\it exactly} and provides 
a simple interpretation of these constraints on $\gamma$. Furthermore, 
it allows us to investigate the sensitivity both on rescattering and on 
possible $SU(3)$-breaking effects (see (\ref{EE1})--(\ref{EE3})). The latter 
may, among other things, lead to $\omega\not=0^\circ$. 

\begin{figure}
\centerline{
\rotate[r]{
\epsfxsize=9.2truecm
\epsffile{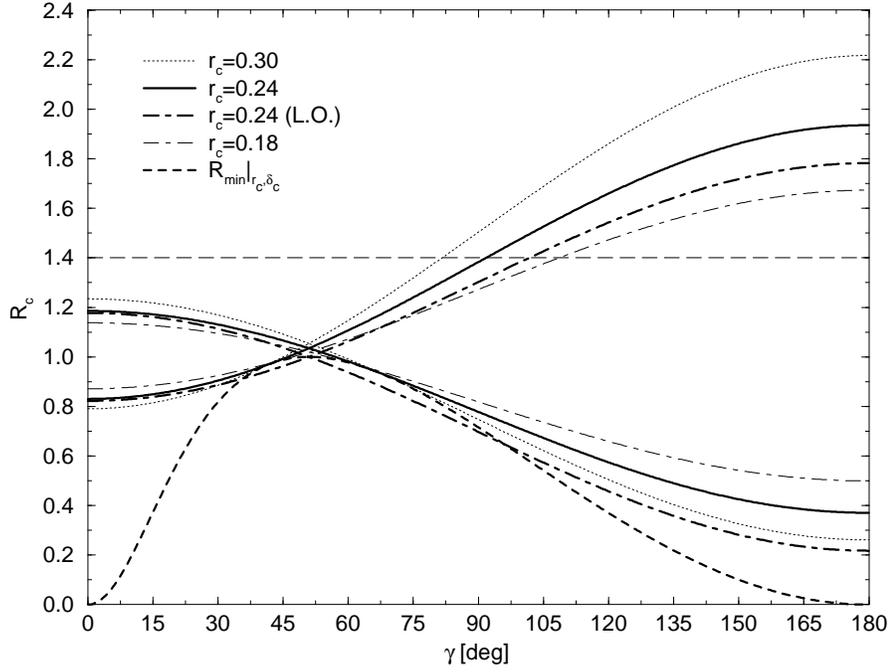}}}
\caption{The dependence of the extremal values of $R_{\rm c}$ described by
(\ref{Rc-bounds}) and (\ref{RcminEWP}) on the CKM angle $\gamma$ for 
$q\,e^{i\omega}=0.63$ in the case of negligible rescattering effects, i.e.\
$\rho=0$.}\label{fig:NR-bounds}
\end{figure}

\begin{figure}
\centerline{
\rotate[r]{
\epsfxsize=9.2truecm
\epsffile{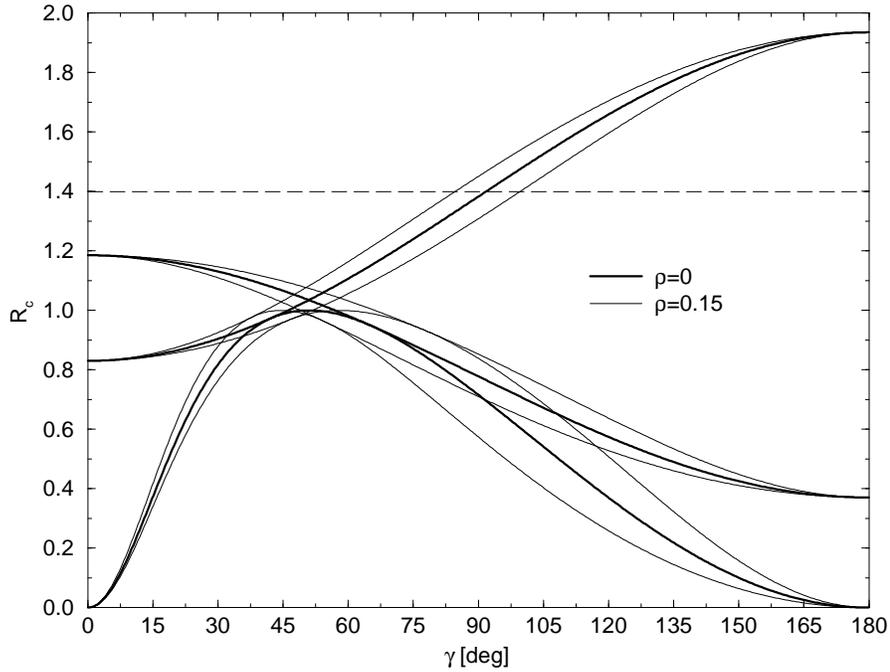}}}
\caption{The impact of rescattering effects on the extremal values of 
$R_{\rm c}$ described by (\ref{Rc-bounds}) and (\ref{RcminEWP})
for $q\,e^{i\omega}=0.63$ and $r_{\rm c}=0.24$ ($\theta\in
\{0^\circ,180^\circ\}$).}\label{fig:RFSI}
\end{figure}

\begin{figure}
\centerline{
\rotate[r]{
\epsfxsize=9.2truecm
\epsffile{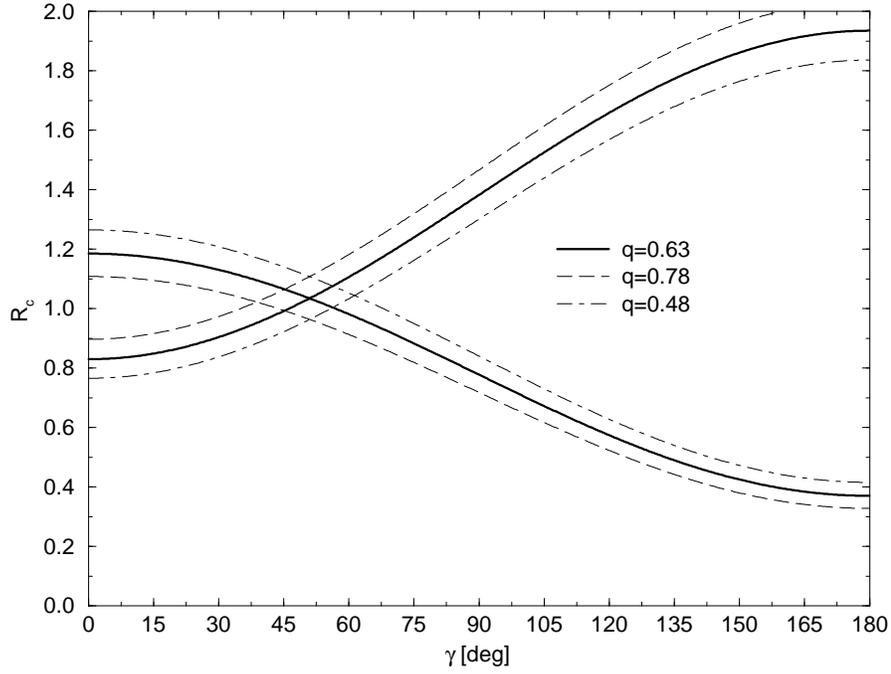}}}
\caption{The dependence of the extremal values of $R_{\rm c}$ 
described by (\ref{Rc-bounds}) on the CKM angle $\gamma$ for 
$r_{\rm c}=0.24$, $\omega=0^\circ$ and for various values of $q$ 
($\rho=0$).}\label{fig:REW}
\end{figure}

\begin{figure}
\centerline{
\rotate[r]{
\epsfxsize=9.2truecm
\epsffile{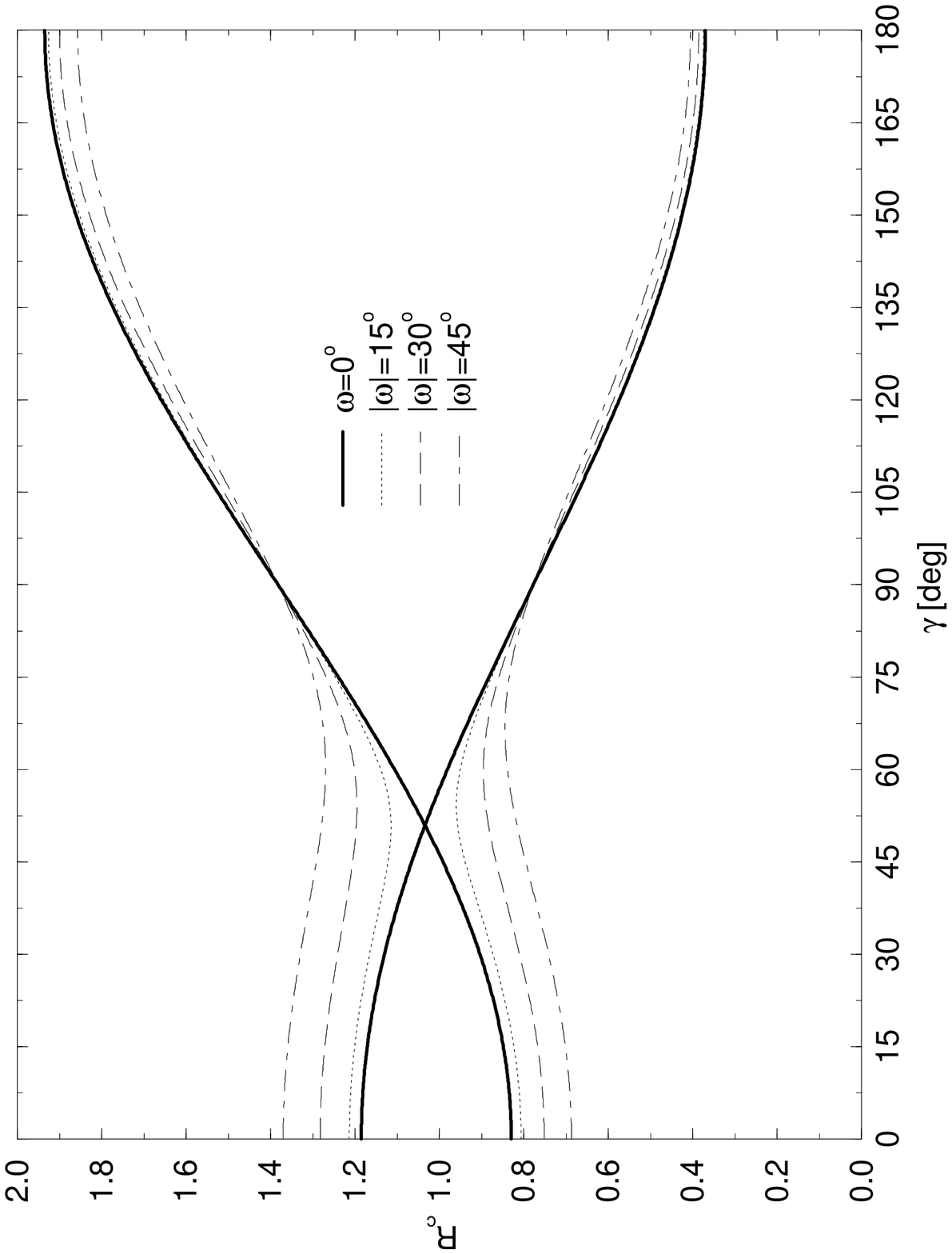}}}
\caption{The dependence of the extremal values of $R_{\rm c}$ described by 
(\ref{Rc-bounds}) on the CKM angle $\gamma$ for $r_{\rm c}=0.24$, $q=0.63$ 
and for various values of $\omega$ ($\rho=0$).}\label{fig:REWphase}
\end{figure}

In Fig.\ \ref{fig:NR-bounds}, we have chosen $q\,e^{i\omega}=0.63$  
to illustrate the constraints on the CKM angle $\gamma$ that are implied
by (\ref{Rc-bounds}) and (\ref{RcminEWP}) for values of $r_{\rm c}$ 
lying within the presently allowed range given in \cite{nr1}. The thick 
dot-dashed line corresponds to the leading-order term of the expansion 
employed by Neubert and Rosner in \cite{nr1}. In the case of $r_{\rm c}=
0.24$ and $R_{\rm c}=1.4$, values of $\gamma<92^\circ$ would be excluded, 
as can be read off easily from this figure. We observe that the lower
bound on $\gamma$ following from the leading-order result given in 
\cite{nr1} receives a sizeable correction of $-\,10^\circ$ in this example.

The extraction of the parameter $r_{\rm c}$ is -- in contrast to the
determination of $r$ in the $B^\pm\to\pi^\pm K$, $B_d\to\pi^\mp K^\pm$ 
case -- not affected by rescattering processes and can be accomplished 
by using only $SU(3)$ flavour symmetry arguments. However, this feature 
does not imply that the constraints on $\gamma$ are also not affected 
by rescattering processes, which may lead to sizeable values of $\rho$.
We have illustrated these effects in Fig.\ \ref{fig:RFSI}, where we have
chosen $q\,e^{i\omega}=0.63$, $r_{\rm c}=0.24$, $\rho=0.15$ and $\theta\in
\{0^\circ,180^\circ\}$. For these strong phases, the rescattering effects
are maximal. In the case of $R_{\rm c}=1.4$, they lead to an uncertainty of
$\Delta\gamma=\pm\,7^\circ$. If we compare these effects with the analysis 
performed in \cite{defan}, we observe that the constraints on $\gamma$ are 
affected, to a similar extent, by rescattering processes, as are those implied 
by the $B^\pm\to\pi^\pm K$ and $B_d\to\pi^\mp K^\pm$ observables \cite{fm2}. 
In our formalism, these effects can be taken into account with the help 
of the strategies proposed in \cite{defan,rf-FSI} (for alternative
methods, see \cite{fknp,bfm}). To this end, additional experimental data 
provided by $B^\pm\to K^\pm K$ decays are needed. Since this issue was 
discussed extensively in \cite{defan,rf-FSI}, we will not work it out in 
more detail here. 

Let us now investigate the uncertainties associated with the electroweak
penguin parameter $q\,e^{i\omega}$. In Fig.\ \ref{fig:REW}, we show
the dependence of (\ref{Rc-bounds}) on the CKM angle $\gamma$ for 
$r_{\rm c}=0.24$, $\omega=0^\circ$ and for various values of $q$. The
strong phase $\omega$ is varied in Fig.~\ref{fig:REWphase} by keeping 
$r_{\rm c}=0.24$ and $q=0.63$ fixed. If we look at these figures, we
observe that in particular non-vanishing values of $\omega$, which may
be induced by non-factorizable $SU(3)$-breaking effects, may weaken 
the bounds on $\gamma$ implied by (\ref{Rc-bounds}) significantly for
$1.2\mathrel{\hbox{\rlap{\hbox{\lower4pt\hbox{$\sim$}}}\hbox{$<$}}}R_{\rm c}
\mathrel{\hbox{\rlap{\hbox{\lower4pt\hbox{$\sim$}}}\hbox{$<$}}}1.4$. As 
we will see in the next subsection, a similar comment applies to the
strategies to determine the CKM angle $\gamma$ with the help of the 
decays $B^\pm\to\pi^\pm K$, $\pi^0K^\pm$. 

\boldmath
\subsection{Strategies to Determine the CKM Angle $\gamma$}
\unboldmath
In analogy to the $B^\pm\to\pi^\pm K$, $B_d\to\pi^\mp K^\pm$ strategy, it is
possible to go beyond the bounds on $\gamma$ discussed in the previous 
subsection as soon as CP violation in $B^\pm\to\pi^0K^\pm$ decays is 
observed. We then are in a position to determine contours in the 
$\gamma$--$r_{\rm c}$ plane with the help of the general formulae given in 
\cite{defan}. Since $r_{\rm c}$ can be fixed through (\ref{T-C-det}), 
$\gamma$ can be determined from these contours, which correspond to a 
mathematical implementation of the simple triangle construction proposed in 
\cite{grl}. However, in contrast to this construction, these contours take 
into account electroweak penguins through (\ref{EE3}).

\begin{figure}
\centerline{
\rotate[r]{
\epsfxsize=9.2truecm
\epsffile{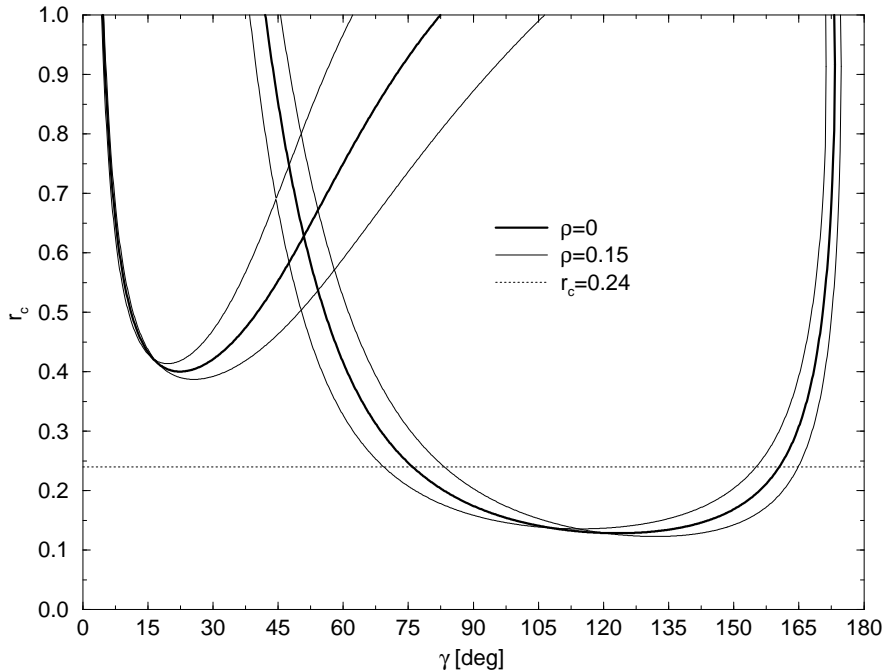}}}
\caption{The contours in the $\gamma$--$r_{\rm c}$ plane corresponding to 
$R_{\rm c}=1.24$, $A_0^{\rm c}=15.9\%$ and $q\,e^{i\omega}=0.63$. The 
thin lines illustrate rescattering effects ($\rho=0.15$, $\theta\in\{0^\circ,
180^\circ\}$).}\label{fig:contours-FSI}
\end{figure}

\begin{figure}
\centerline{
\rotate[r]{
\epsfxsize=9.2truecm
\epsffile{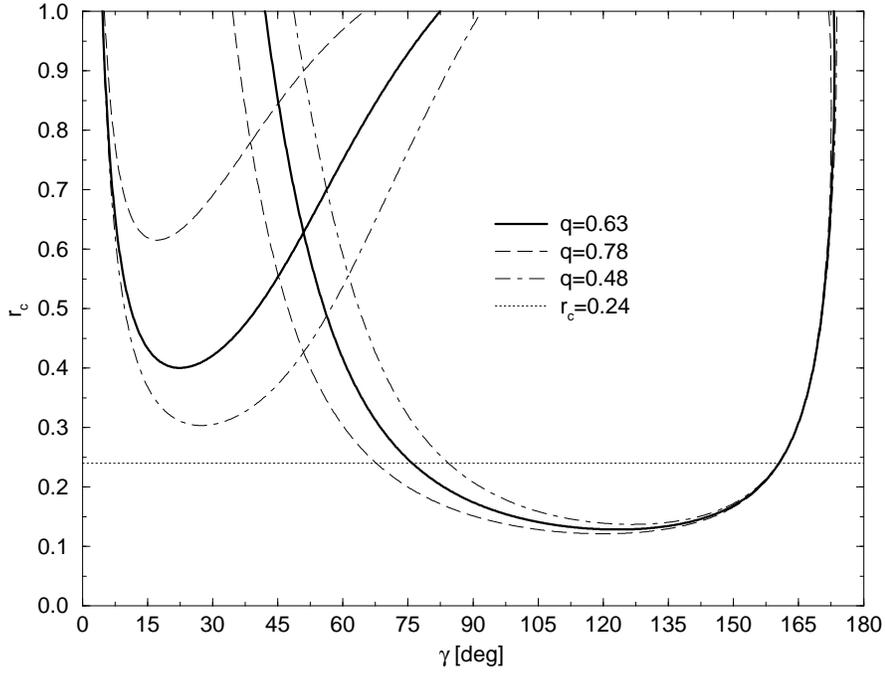}}}
\caption{The dependence of the contours in the $\gamma$--$r_{\rm c}$ plane 
corresponding to $R_{\rm c}=1.24$, $A_0^{\rm c}=15.9\%$ and $\omega=0^\circ$
on the parameter $q$ ($\rho=0$).}\label{fig:contours-EWP2}
\end{figure}

\begin{figure}
\centerline{
\rotate[r]{
\epsfxsize=9.2truecm
\epsffile{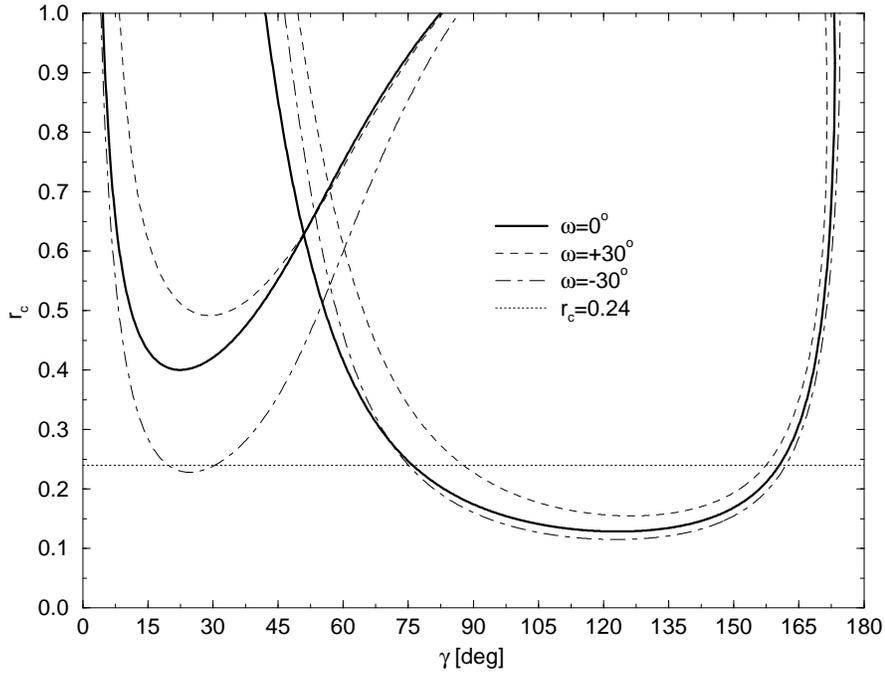}}}
\caption{The impact of non-vanishing values of the strong phase $\omega$ 
on the contours in the $\gamma$--$r_{\rm c}$ plane corresponding to 
$R_{\rm c}=1.24$, $A_0^{\rm c}=15.9\%$ and $q=0.63$ 
($\rho=0$).}\label{fig:contours-EWP1}
\end{figure}

Let us consider again a specific example in order to illustrate this 
strategy in more detail. To this end, we follow Neubert and Rosner 
\cite{nr2} and assume that $\gamma=76^\circ$, $\rho=0$, $r_{\rm c}=0.24$, 
$\delta_{\rm c}=20^\circ$ and $q\,e^{i\omega}=0.63$ to calculate the 
observables $R_{\rm c}$ and $A_0^{\rm c}$. These parameters give 
$R_{\rm c}=1.24$ and $A_0^{\rm c}=15.9\%$. In Fig.\ \ref{fig:contours-FSI}, 
we show the corresponding contours in the $\gamma$--$r_{\rm c}$ plane. 
The thick lines describe the contours 
arising for $\rho=0$, and the dotted line represents the ``measured''
value of $r_{\rm c}$. Their intersection gives a two-fold solution for
$\gamma$, including the ``true'' value of $76^\circ$ and a second solution
of $160^\circ$. The thin lines illustrate the impact of possible rescattering 
processes and are obtained for $\rho=0.15$ and $\theta\in\{0^\circ,
180^\circ\}$. For these strong phases, the rescattering effects are maximal. 
Applying the strategies proposed in \cite{defan,rf-FSI}, the rescattering 
effects can be taken into account in these contours. To this end,
additional experimental data on $B^\pm\to K^\pm K$ decays are required. 
Unfortunately, non-factorizable $SU(3)$-breaking effects cannot be included
in a similar manner. Let us emphasize again at this point that such 
corrections may affect both the determination of $|T+C|$, i.e.\ of 
$r_{\rm c}$, and the calculation of the electroweak penguin parameter 
$q\,e^{i\omega}$, which may therefore be shifted significantly from 
(\ref{EE3}). In Fig.\ \ref{fig:contours-EWP2}, we show the dependence of the 
contours in the $\gamma$--$r_{\rm c}$ plane arising in our specific example 
on the parameter $q$, while we illustrate the impact of non-vanishing values 
of the strong phase $\omega$ in Fig.\ \ref{fig:contours-EWP1}. In these two
figures, we have neglected rescattering effects, i.e.\ we have chosen 
$\rho=0$. We observe that the contours are rather sensitive to the phase 
$\omega$. For values of $\omega=-\,30^\circ$, we even get additional 
solutions for $\gamma$.

Before we present a new strategy to probe the CKM angle $\gamma$ with the
help of the neutral decays $B_d\to\pi^0K$, $\pi^\mp K^\pm$, let us briefly 
go back to the $B^\pm\to\pi^\pm K$, $B_d\to\pi^\mp K^\pm$ approach
discussed in Section~\ref{PAPIII-strat}. If we compare the contours in
the $\gamma$--$r$ plane shown in Fig.~\ref{fig:FM-contours} with those
in the $\gamma$--$r_{\rm c}$ plane shown in Fig.\ 
\ref{fig:contours-FSI}, we observe that they are very different from 
each other. In particular, the $B^\pm\to\pi^\pm K$, $B_d\to\pi^\mp K^\pm$ 
case appears to be more promising for this specific example. Time will tell 
which one of these two strategies is really more powerful in practice.

\boldmath
\section{Probing the CKM Angle $\gamma$ with the Neutral\\
Decays $B_d\to\pi^0K$ and $B_d\to\pi^\mp K^\pm$}\label{neut-strat}
\unboldmath 
The observables $R_{\rm n}$ and $A_0^{\rm n}$ of the neutral $B$ decays
$B_d\to\pi^0K$, $\pi^\mp K^\pm$ allow strategies to probe the CKM 
angle $\gamma$ that are completely analogous to those discussed in the
previous section. However, in the case of these modes, we have an additional 
CP-violating observable at our disposal, which allows us to take into account 
rescattering effects in a theoretically clean way. The point is as follows: 
since $B_d\to\pi^\mp K^\pm$ is a self-tagging neutral $B$ decay, it exhibits 
only direct CP violation due to the interference between the ``tree'' 
and ``penguin'' amplitudes, but no mixing-induced CP violation, 
arising from interference effects between $B^0_d$--$\overline{B^0_d}$ mixing 
and decay processes. On the other hand, if we consider $B_d\to\pi^0K$ modes
and require that the kaon be observed as a $K_{\rm S}$, the resulting final 
state $f$ is an eigenstate of the CP operator with eigenvalue $-1$. In this 
case, we have to deal with mixing-induced CP violation and obtain the 
following time-dependent CP asymmetry \cite{rev}:
\begin{eqnarray}
\lefteqn{A_{\rm CP}(B_d(t)\to f)\equiv\frac{\Gamma(B_d^0(t)\to f)\,-\,
\Gamma(\overline{B^0_d}(t)\to f)}{\Gamma(B_d^0(t)\to f)\,+
\,\Gamma(\overline{B^0_d}(t)\to f)}}\nonumber\\
&=&{\cal A}_{\rm CP}^{\rm dir}(B_d\to f)\,\cos(\Delta M_d\,t)\,+\,
{\cal A}_{\rm CP}^{\rm mix-ind}(B_d\to f)\,\sin(\Delta M_d\,t)\,.\label{CPASY}
\end{eqnarray}
Here $\Gamma(B_d^0(t)\to f)$ and $\Gamma(\overline{B^0_d}(t)\to f)$ denote 
the decay rates of initially, i.e.\ at time $t=0$, present $B_d^0$ and 
$\overline{B^0_d}$ mesons, respectively; $\Delta M_d$ is the mass 
difference of the $B_d$ mass eigenstates, and 
\begin{equation}\label{ACP-dir}
{\cal A}_{\rm CP}^{\rm dir}(B_d\to f)=\frac{1-\left|\xi_f^{(d)}\right|^2}{1+
\left|\xi_f^{(d)}\right|^2}
\end{equation}
\begin{equation}\label{ACP-mix}
{\cal A}_{\rm CP}^{\rm mix-ind}(B_d\to f)=
\frac{2\,\mbox{Im}\,\xi_f^{(d)}}{1+\left|\xi_f^{(d)}\right|^2}
\end{equation}
describe direct and mixing-induced CP violation. The observable 
$\xi_f^{(d)}$ containing essentially all the information needed to evaluate 
these CP-violating asymmetries is given as follows:
\begin{equation}\label{xi-def}
\xi_f^{(d)}=\mp\,e^{-i \phi_{\rm M}^{(d)}}\frac{A(\overline{B^0_d}\to 
f)}{A(B^0_d\to f)},
\end{equation}
where $A(B^0_d\to f)$ and $A(\overline{B^0_d}\to f)$ are ``unmixed'' 
decay amplitudes, $\phi_{\rm M}^{(d)}=2\,\mbox{arg}(V_{td}^\ast V_{tb})$
denotes the weak $B^0_d$--$\overline{B^0_d}$ mixing phase, and $({\cal CP})
|f\rangle=\pm|f\rangle$. If the final state $f$ contains a neutral kaon, 
as in the case of $B_d\to\pi^0K_{\rm S}$, we have in addition to take into 
account a phase $\phi_K$, which is related to $K^0$--$\overline{K^0}$ mixing 
and is negligibly small in the Standard Model. The combination 
$\phi_{\rm M}^{(d)}+\phi_K$, which is relevant for $B_d\to\pi^0K_{\rm S}$, 
can be determined in a theoretically clean way with the help of 
the ``gold-plated'' mode $B_d\to J/\psi\,K_{\rm S}$ \cite{csbs}:
\begin{equation}\label{psiKS}
A_{\rm CP}(B_d(t)\to J/\psi\,K_{\rm S})=-\,\sin\left(\phi_{\rm M}^{(d)}+
\phi_K\right)\,\sin(\Delta M_d\,t)\,.
\end{equation}
Within the Standard Model, we have $\phi_{\rm M}^{(d)}=2\beta$, where 
$\beta$ is another angle of the unitarity triangle, and $\phi_K=0$ to a
very good approximation. In the case of the decay $B_d\to\pi^0K_{\rm S}$, 
the observables (\ref{ACP-dir}) and (\ref{ACP-mix}) can be expressed as 
follows \cite{PAPII,bsgam}:
\begin{eqnarray}
{\cal A}_{\rm CP}^{\rm dir}(B_d\to\pi^0K_{\rm S})&=&
\frac{|P_{\rm n}|^2-|\overline{P_{\rm n}}|^2}{|P_{\rm n}|^2+
|\overline{P_{\rm n}}|^2}\label{dir-exp}\\
{\cal A}_{\rm CP}^{\rm mix-ind}(B_d\to\pi^0K_{\rm S})&=&
-\,\frac{2\,|P_{\rm n}||\overline{P_{\rm n}}|}{|P_{\rm n}|^2+
|\overline{P_{\rm n}}|^2}\,\sin\left[\left(\phi_{\rm M}^{(d)}+\phi_K\right)+
\psi\right],\label{mix-exp}
\end{eqnarray}
where $P_{\rm n}\equiv\sqrt{2}\,A(B^0_d\to\pi^0K^0)$ (see (\ref{Bnampl})),
$\overline{P_{\rm n}}\equiv\sqrt{2}\,A(\overline{B^0_d}\to\pi^0
\overline{K^0})$, and $\psi$ denotes the angle between these amplitudes,
i.e.\ $\overline{P_{\rm n}}/P_{\rm n}\equiv e^{-i\psi}\,
|\overline{P_{\rm n}}|/|P_{\rm n}|$.

\begin{figure}
\centerline{
\rotate[r]{
\epsfxsize=6.5truecm
\epsffile{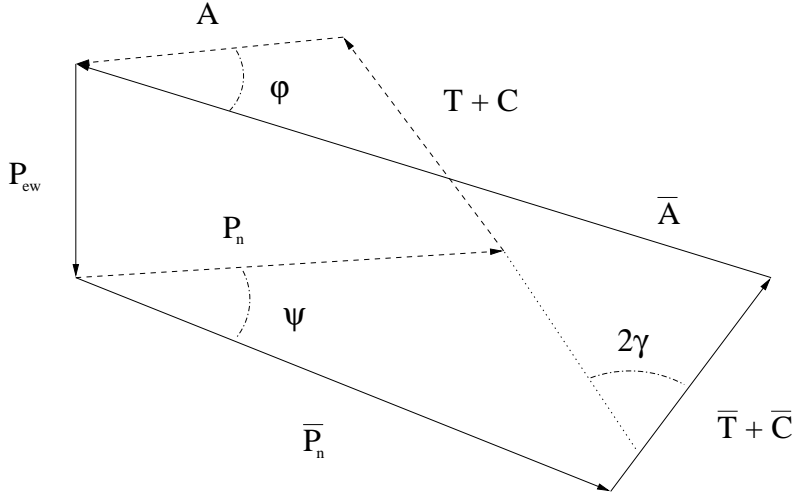}}}
\vspace*{0.7truecm} 
\caption{Illustration of a strategy to determine the CKM angle $\gamma$
by means of the neutral decays $B^0_d\to\pi^0K^0$, $B^0_d\to\pi^-K^+$ and 
their charge conjugates.}\label{fig:constr}
\end{figure}

The determination of $\gamma$ by means of $B_d\to\pi^0K$, $\pi^\mp K^\pm$, 
which we would like to propose here, uses Eqs.\ (\ref{ampl-rel2}), 
(\ref{EE3}), (\ref{T-C-det}), (\ref{CPASY}) and (\ref{psiKS})--(\ref{mix-exp}).
The geometrical version of this determination, which is illustrated in
Fig.\ \ref{fig:constr}, proceeds in the following steps:
\begin{enumerate}
\item From time-dependent studies of the $B_d\to\pi^0K_{\rm S}$ and 
$B_d\to J/\psi\,K_{\rm S}$ decay rates and the associated CP-violating
asymmetries, which are represented by  (\ref{CPASY}) and 
(\ref{psiKS})--(\ref{mix-exp}), we determine the absolute values of
the amplitudes $P_{\rm n}$ and $\overline{P_{\rm n}}$, as well as
their relative orientation in the complex plane, i.e.\ the angle $\psi$.
\item Using (\ref{EE3}) and (\ref{T-C-det}), we determine $|P_{\rm ew}|$
and $|T+C|=|\overline{T}+\overline{C}|$, respectively.
\item Using BR$(B^0_d\to\pi^-K^+)$ and BR$(\overline{B^0_d}\to\pi^+K^-)$, 
we determine the magnitudes of the amplitudes $A\equiv A(B^0_d\to\pi^-K^+)$ 
and $\overline{A}\equiv A(\overline{B^0_d}\to\pi^+K^-)$, respectively.
\item The information collected in steps 1--3 allows us to construct two
quadrangles in the complex plane, as shown in Fig.\ \ref{fig:constr}. They
are a geometrical representation of the amplitude relation (\ref{ampl-rel2}) 
and its CP conjugate, which -- in terms of the notation used in this figure --
take the form
\begin{eqnarray}
P_{\rm n}+(T+C)+A+P_{\rm ew}&=&0\\
\overline{P_{\rm n}}+(\overline{T}+\overline{C})+\overline{A}+P_{\rm ew}&=&0\,.
\end{eqnarray}
Since only information on $|P_{\rm ew}|$ has been used so far, the precise 
shapes of these two quadrangles are not yet fixed.
\item Finally, we make again use of (\ref{EE3}) to determine the 
phase $\omega=\delta_{\rm ew}-\delta_{T+C}$. This gives us the
orientation of the electroweak penguin amplitude $P_{\rm ew}$ with
respect to the line that bisects the angle between $T+C$ and $\overline{T}+
\overline{C}$. This final information, together with the construction of
step 4, determines the shapes of the two quadrangles in question, and
consequently also the CKM angle $\gamma$, as shown in Fig.\ \ref{fig:constr}.
\end{enumerate}
In this construction, there are no uncertainties due to rescattering 
effects, and the theoretical accuracy is limited only by non-factorizable 
$SU(3)$-breaking corrections, which may affect (\ref{EE3}) and 
(\ref{T-C-det}). 

In order to have the tools available to implement this geometrical 
construction in a mathematical way, we give the explicit expression for 
${\cal A}_{\rm CP}^{\rm mix-ind}(B_d\to\pi^0K_{\rm S})$ in terms of the 
parameters $\rho_{\rm n}$ and $\theta_{\rm n}$ defined in (\ref{Bnampl}):
\begin{eqnarray}
\lefteqn{{\cal A}_{\rm CP}^{\rm mix-ind}(B_d\to\pi^0K_{\rm S})=}\\
&&-\left[\frac{\sin\left(\phi_{\rm M}^{(d)}+\phi_K\right)+2\,\rho_{\rm n}
\cos\theta_{\rm n}\sin\left(\phi_{\rm M}^{(d)}+\phi_K+\gamma\right)+
\rho_{\rm n}^2\sin\left(\phi_{\rm M}^{(d)}+\phi_K+2\,\gamma\right)}{1+
2\,\rho_{\rm n}\cos\theta_{\rm n}\cos\gamma+\rho_{\rm n}^2}\right],\nonumber
\end{eqnarray}
which reduces to 
\begin{equation}\label{CP-rel}
{\cal A}_{\rm CP}^{\rm mix-ind}(B_d\to\pi^0K_{\rm S})=-\,
\sin\left(\phi_{\rm M}^{(d)}+\phi_K\right)=
{\cal A}_{\rm CP}^{\rm mix-ind}(B_d\to J/\psi\,K_{\rm S})
\end{equation} 
in the case of $\rho_{\rm n}=0$ \cite{PAPIII}. The direct CP asymmetry 
${\cal A}_{\rm CP}^{\rm dir}(B_d\to\pi^0K_{\rm S})$ takes the same form 
as the direct CP asymmetry $A_+$ arising in the decay $B^+\to\pi^+K^0$ 
(see (\ref{Ap-def})). Consequently, measuring ${\cal A}_{\rm CP}^{\rm dir}
(B_d\to\pi^0K_{\rm S})$, ${\cal A}_{\rm CP}^{\rm mix-ind}
(B_d\to\pi^0K_{\rm S})$, $R_{\rm n}$ and $A_0^{\rm n}$ (see (\ref{Rn-def}) 
and (\ref{A0n-def})), we can determine the four ``unknowns'' $\rho_{\rm n}$, 
$\theta_{\rm n}$, $\delta_{\rm n}$ and the CKM angle $\gamma$ ($r_{\rm n}$ 
and $\phi_{\rm M}^{(d)}+\phi_K$ are fixed through (\ref{T-C-det}) and 
(\ref{psiKS}), respectively) as functions of the electroweak penguin parameter 
$q\,e^{i\omega}$. The latter can be determined by using (\ref{EE3}).

The utility of time-dependent measurements of the decay $B_d\to\pi^0K_{\rm S}$
to probe angles of the unitarity triangle was already pointed out several
years ago by Nir and Quinn \cite{nq}, who proposed a strategy to determine
the angle $\alpha$ with the help of the amplitude relation (\ref{ampl-rel2}).
At that time, it was believed that electroweak penguins play only a very
minor role in $B$ decays, which is actually not the case because of the 
large top-quark mass \cite{rf-ewp,rf-ewp3}. A construction similar to the 
one shown in Fig.\ \ref{fig:constr} would in fact allow the extraction of 
the CKM angle $\alpha$, if electroweak penguins played a negligible role, 
i.e.\ if $P_{\rm ew}=0$. In order to see how this strategy works, we have
to rotate the CP-conjugate amplitudes $\overline{P_{\rm n}}$, 
$\overline{T}+\overline{C}$ and $\overline{A}$ by the phase factor 
$e^{-i(\phi_{\rm M}^{(d)}+\phi_K)}=e^{-i2\beta}$, so that the 
angle between $T+C$ and the rotated $\overline{T}+\overline{C}$ amplitude 
is a measure of $2\alpha$ (note that $\beta+\gamma=180^\circ-\alpha$). 
A similar ``trick'' was also used in our discussion of the $B\to\pi\pi$ 
approach given in the appendix. Since the angle 
$\tilde\psi\equiv(\phi_{\rm M}^{(d)}+\phi_K)+\psi$ between $P_{\rm n}$ 
and the rotated amplitude $\overline{P_{\rm n}}$ can be determined directly 
by using the mixing-induced CP asymmetry (\ref{mix-exp}), the CKM angle 
$\alpha$ can be determined. Unfortunately, this construction does not work
in the presence of electroweak penguins. In order to take them into account
with the help of (\ref{EE3}), the phase $\phi_{\rm M}^{(d)}+\phi_K$ of the 
rotated electroweak penguin amplitude $\overline{P_{\rm ew}}(=P_{\rm ew})$ 
has to be determined by making use of (\ref{psiKS}), and we arrive at a 
construction, which is equivalent to the one discussed above. Interestingly, 
the situation in this respect is very different in the $\alpha$ determination 
from $B\to\pi\pi$ isospin triangle relations, as we show in the appendix.

Let us now come back to the decay $B_d\to\pi^0K$. Concerning the parameter 
$\rho_{\rm n}$ defined in (\ref{rho-n-def}), the usual na\"\i ve expectation 
based on ``colour suppression'' and ``short-distance'' arguments is a value 
at the level of a few per cent, implying small direct CP violation in 
$B_d\to\pi^0K$ and small corrections to (\ref{CP-rel}). Moreover, we would 
expect a tiny angle $\psi$ between the amplitudes $P_{\rm n}$ and 
$\overline{P_{\rm n}}$ in Fig.\ \ref{fig:constr}. However, rescattering 
effects of the kind discussed in \cite{neubert}, \cite{FSI}--\cite{groro-FSI} 
may in principle also lead to an enhancement of $\rho_{\rm n}$, thereby 
affecting (\ref{CP-rel}) and leading to sizeable direct CP violation in 
$B_d\to\pi^0K$, as well as to a sizeable value of the angle $\psi$. On the 
other hand, if (\ref{dir-exp}) and (\ref{mix-exp}) should in fact imply a 
tiny value of $\psi$, i.e.\ that $P_{\rm n}\approx\overline{P_{\rm n}}$, 
there would be a simple strategy to extract $\gamma$ by using in addition 
the observables provided by an analysis of the decay $B_s\to K^+K^-$. For 
sizeable values of $\psi$, this mode would also be very useful, allowing 
us to reduce the theoretical input concerning the electroweak penguins
considerably. Let us turn to this decay in the following section. 

\boldmath
\section{Strategies to Combine $B_s\to K^+K^-$ Modes with 
$B_{u,d}\to\pi K$ Decays}\label{BsKK}
\unboldmath
\subsection{Preliminaries}
The decay $B_s\to K^+K^-$, which is the $B_s$ counterpart of the mode
$B_d\to\pi^\mp K^\pm$, plays an important role to probe the CKM angle
$\gamma$ and to obtain experimental insights into electroweak 
penguins [3, 17, 39--41]. In contrast to the $B_d$ case, 
there may be a sizeable width difference $\Delta\Gamma_s\equiv
\Gamma_{\rm H}^{(s)}-\Gamma_{\rm L}^{(s)}$ between the mass eigenstates 
$B_s^{\rm H}$ (``heavy'') and $B_s^{\rm L}$ (``light'') of the $B_s$ 
system \cite{DGamma}, which may allow studies of CP violation with untagged
$B_s$ data samples, where one does not distinguish between initially, i.e.\ 
at time $t=0$, present $B_s^0$ or $\overline{B^0_s}$ mesons \cite{dunietz}.
The corresponding untagged $B_s$ decay rates are defined by 
\begin{equation}
\Gamma[f(t)]\equiv\Gamma(B_s^0(t)\to f)+\Gamma(\overline{B^0_s}(t)\to f),
\end{equation}
and can be expressed as (see, for instance, \cite{rev})
\begin{equation}\label{untagged}
\Gamma[f(t)]\propto \Bigl[1+{\cal A}_{\Delta\Gamma}(B_s\to f)\Bigr]\,
e^{-\Gamma_{\rm H}^{(s)}t}+\Bigl[1-{\cal A}_{\Delta\Gamma}(B_s\to f)\Bigr]\,
e^{-\Gamma_{\rm L}^{(s)}t}
\end{equation}
with
\begin{equation}
{\cal A}_{\Delta\Gamma}(B_s\to f)=
\frac{2\,\mbox{Re}\,\xi_f^{(s)}}{1+\left|\xi_f^{(s)}\right|^2}.
\end{equation}
Note that there are no rapid oscillatory $\Delta M_s\,t$ terms present 
in (\ref{untagged}). The observable $\xi_f^{(s)}$ is defined in analogy 
to (\ref{xi-def}); we have just to replace the $B_d^0$--$\overline{B^0_d}$ 
mixing phase $\phi_{\rm M}^{(d)}$ in that expression by its $B_s$ counterpart 
$\phi_{\rm M}^{(s)}=2\,\mbox{arg}(V_{ts}^\ast V_{tb})$, which is negligibly 
small in the Standard Model. The width difference $\Delta\Gamma_s$ modifies 
also the expression for the time-dependent CP asymmetry (\ref{CPASY}). In 
the $B_s$ case, it takes the following form:
\begin{eqnarray}
\lefteqn{A_{\rm CP}(B_s(t)\to f)\equiv\frac{\Gamma(B_s^0(t)\to f)\,-\,
\Gamma(\overline{B^0_s}(t)\to f)}{\Gamma(B_s^0(t)\to f)\,+
\,\Gamma(\overline{B^0_s}(t)\to f)}}\nonumber\\
&=&2\,e^{-\Gamma_st}\left[\frac{{\cal A}_{\rm CP}^{\rm
dir}(B_s\to f)\cos(\Delta M_s\,t)+{\cal A}_{\rm CP}^{\rm mix-ind}(B_s\to f)
\sin(\Delta M_s\,t)}{e^{-\Gamma_{\rm H}^{(s)}t}+e^{-\Gamma_{\rm L}^{(s)}t}
\,+\,{\cal A}_{\Delta\Gamma}(B_s\to f)
\left(e^{-\Gamma_{\rm H}^{(s)}t}\,-\,e^{-\Gamma_{\rm L}^{(s)}t}\right)}\right],
\end{eqnarray}
where ${\cal A}_{\rm CP}^{\rm dir}(B_s\to f)$ and ${\cal A}_{\rm 
CP}^{\rm mix-ind}(B_s\to f)$ correspond to (\ref{ACP-dir}) and 
(\ref{ACP-mix}), respectively, and $\Gamma_s\equiv\left(\Gamma_{\rm H}^{(s)}+
\Gamma_{\rm L}^{(s)}\right)/2$.

If we introduce the notation $A_s\equiv A(B_s^0\to K^+K^-)$, $\overline{A_s}
\equiv A(\overline{B_s^0}\to K^+K^-)$ and denote the angle between these
amplitudes by $\varphi_s$, we obtain the following expressions for the
$B_s\to K^+K^-$ observables \cite{bsgam}:
\begin{eqnarray}
{\cal A}_{\rm CP}^{\rm dir}(B_s\to K^+K^-)&=&
\frac{|A_s|^2-|\overline{A_s}|^2}{|A_s|^2+|\overline{A_s}|^2}\label{Bs-dir}\\
{\cal A}_{\rm CP}^{\rm mix-ind}(B_s\to K^+K^-)&=&
\frac{2\,|A_s||\overline{A_s}|}{|A_s|^2+|\overline{A_s}|^2}\,\sin
\left(\phi_{\rm M}^{(s)}+\varphi_s\right)\label{Bs-mix}\\
{\cal A}_{\Delta\Gamma}(B_s\to K^+K^-)&=&
-\,\frac{2\,|A_s||\overline{A_s}|}{|A_s|^2+|\overline{A_s}|^2}\,\cos
\left(\phi_{\rm M}^{(s)}+\varphi_s\right)\label{Bs-DG}.
\end{eqnarray}
The measurement of these quantities allows us to construct the amplitudes
$A_s$ and $\overline{A_s}$ in the complex plane, i.e.\ to determine both
their magnitudes and their relative orientation, provided the 
$B^0_s$--$\overline{B^0_s}$ mixing phase $\phi_{\rm M}^{(s)}$ is known. 
As we already noted, this phase is tiny in the Standard Model. 
It can in principle be determined with the help of the decay 
$B_s\to J/\psi\,\phi$ (see, for example, \cite{fd1,ddf1}), which is the 
$B_s$ counterpart of the ``gold-plated'' mode $B_d\to J/\psi\,K_{\rm S}$ 
and is very accessible at future hadron machines, for example at the LHC. 
Large CP violation in $B_s\to J/\psi\,\phi$ would indicate new-physics 
contributions to $B^0_s$--$\overline{B^0_s}$ mixing. Even in such a scenario 
of new physics, it would be possible to fix the amplitudes $A_s$ and 
$\overline{A_s}$ in the complex plane by measuring in addition to 
(\ref{Bs-dir})--(\ref{Bs-DG}) the observables of the decay 
$B_s\to J/\psi\,\phi$. 

The decays $B_s\to K^+K^-$ and $B_d\to\pi^\mp K^\pm$ differ only in their
``spectator'' quarks and can be related to each other through $SU(3)$ 
flavour symmetry arguments. Potential $SU(3)$-breaking effects are
also due to ``penguin annihilation'' processes, which contribute to 
$B_s\to K^+K^-$ (for an explicit expression of the decay amplitude, see
\cite{bsgam}), and are absent in $B_d\to\pi^\mp K^\pm$. The importance of 
these topologies, which are expected to play a minor role
\cite{ghlr,ghlr2}, can be investigated with the help of the decay 
$B_s\to\pi^+\pi^-$; other interesting probes for $SU(3)$-breaking effects 
can be obtained by comparing the observables of the untagged $B_s\to K^+K^-$ 
rate with the combined $B_d\to\pi^\mp K^\pm$ branching ratio, or their direct 
CP asymmetries \cite{bsgam}. Let us assume in the following that explorations 
of this kind indicate small $SU(3)$-breaking effects. Then we may identify 
the angle $\varphi_s$ between the $B_s\to K^+K^-$ amplitudes $A_s$ and 
$\overline{A_s}$ with the angle $\varphi$ between the $B_d\to\pi^\mp K^\pm$ 
amplitudes $A$ and $\overline{A}$ (see Fig.\ \ref{fig:constr}). The knowledge 
of this angle would be very useful, since it allows us to fix the relative
orientation of $A$ and $\overline{A}$.

Let us note that a time-dependent, tagged $B_s\to K^+K^-$ analysis has 
to be performed in order to determine $\varphi_s$. However, if we use 
the direct CP asymmetry $A_{\rm CP}(B_d\to\pi^\mp K^\pm)$, the ratio 
$|\overline{A_s}|/|A_s|$ can be fixed with the help of the $SU(3)$ flavour 
symmetry, allowing the determination of $\varphi_s$ up to a two-fold 
ambiguity from the untagged observable (\ref{Bs-DG}). Although future 
$B$-physics experiments performed at hadron machines should be in a 
position to resolve the rapid oscillatory $\Delta M_s\,t$ terms arising 
in tagged $B_s$ data samples, untagged studies are more promising in terms
of efficiency, acceptance and purity \cite{dunietz}.

\subsection{Strategy A}
If the angle $\varphi$ in Fig.\ \ref{fig:constr} is known, the 
theoretical input concerning the electroweak penguin amplitude $P_{\rm ew}$
can be reduced considerably. In particular, step 5 of the procedure 
given in the previous section could be avoided, since the first four 
steps, together with the knowledge of the angle $\varphi$, would
determine the shapes of the two quadrangles in Fig.\ \ref{fig:constr}.
This would not only allow us to determine the CKM angle $\gamma$, 
but also the strong phase $\omega$ in (\ref{EE3}). Conversely, we could
use $\omega$ as our theoretical input to deal with the electroweak 
penguins, and could then determine both the electroweak penguin parameter
$q$ and the CKM angle $\gamma$. Both approaches would offer some
consistency checks for (\ref{EE3}).

Should it become possible to determine the CKM angle $\gamma$ with the 
help of other strategies, using for example the theoretically clean approach
provided by the ``tree'' decays $B_s\to D_s^\pm K^\mp$ \cite{adk}, the
geometrical construction shown in Fig.\ \ref{fig:constr} would allow us
to determine the electroweak penguin amplitude $P_{\rm ew}$ completely,
i.e.\ both $q$ and $\omega$ (see also \cite{PAPI}). To accomplish this task, 
a sizeable angle $\psi$ between the amplitudes $P_{\rm n}$ and 
$\overline{P_{\rm n}}$ is required. Consequently, this strategy to 
determine the electroweak penguin amplitude does not work in the case 
of small rescattering effects and significant ``colour suppression'' 
in $B_d\to\pi^0K$, leading to $P_{\rm n}\approx\overline{P_{\rm n}}$. 
As we will see in the next subsection, there is, however, another, simpler 
strategy to obtain insights into electroweak penguins in this case. 

\subsection{Strategy B}
\begin{figure}
\centerline{
\rotate[r]{
\epsfxsize=4truecm
\epsffile{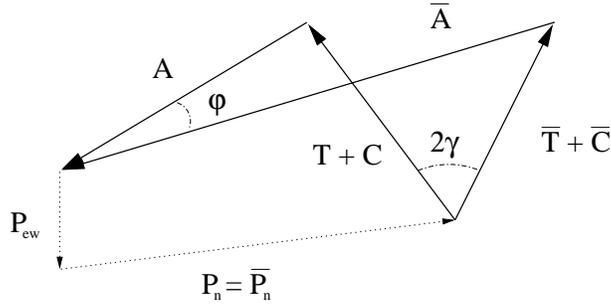}}}
\vspace*{0.7truecm} 
\caption{Simple strategy to determine $\gamma$ with the help of the decays
$B^\pm\to\pi^\pm \pi^0$, $B_d\to\pi^\mp K^\pm$ and $B_s\to K^+K^-$ in the
case of $P_{\rm n}=\overline{P_{\rm n}}$ (thick solid lines), and to obtain 
insights into electroweak penguins by using in addition $B_d\to\pi^0K$ (thin 
dotted lines).}\label{fig:simple-constr}
\end{figure}

The case $P_{\rm n}\approx\overline{P_{\rm n}}$ would be very favourable for 
the extraction of $\gamma$, thereby offering a new way to determine this 
angle that is only affected to a small extent by electroweak penguins. For 
$P_{\rm n}=\overline{P_{\rm n}}$, there would be no 
electroweak penguin uncertainties at all. This strategy requires only the 
measurement of $B^+\to\pi^+\pi^0$ to fix $|T+C|$ with the help of 
(\ref{T-C-det}), and analyses of the decays $B_d\to\pi^\mp K^\pm$ and 
$B_s\to K^+K^-$ to determine the amplitudes $A$ and $\overline{A}$ in the 
complex plane. Although it is possible to see already in 
Fig.~\ref{fig:constr} how this $SU(3)$ strategy works, we think it 
useful to redraw it for the special case of $P_{\rm n}=\overline{P_{\rm n}}$ 
in Fig.\ \ref{fig:simple-constr}. Here the CKM angle $\gamma$ can be 
determined with the help of the simple geometrical construction involving 
only the thick solid lines. The $B_d\to\pi^0 K$ amplitude allows us, 
furthermore, to fix the electroweak penguin parameter $q$, if $\omega$ is 
used as an input, or the strong phase $\omega$, if we use $q$ as an input,
thereby providing consistency checks for (\ref{EE3}).

\subsection{Strategy C}
\begin{figure}
\centerline{
\rotate[r]{
\epsfxsize=3.3truecm
\epsffile{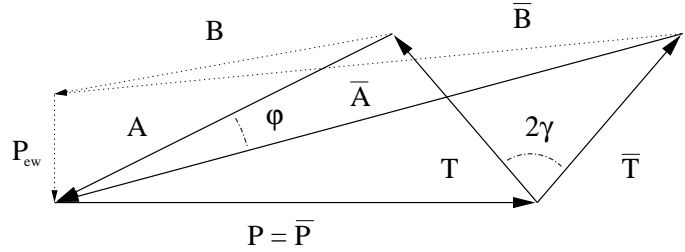}}}
\vspace*{0.7truecm} 
\caption{Simple strategy to determine $\gamma$ with the help of the decays
$B_d\to\pi^\mp K^\pm$ and $B^\pm\to\pi^\pm K$ (thick solid lines), and
to obtain insights into electroweak penguins by using in addition 
$B^\pm\to\pi^0K^\pm$ (thin dotted lines). Here rescattering effects have
been neglected and it has been assumed that ``colour suppression'' is 
effective.}\label{fig:EW-det}
\end{figure}

The $B_d\to\pi^\mp K^\pm$ amplitudes $A$ and $\overline{A}$ can also be 
combined with those of the charged $B$-meson decays $B^\pm\to\pi^\pm K$, 
$\pi^0K^\pm$. Neglecting $P_{\rm ew}^{\rm C}$ in the amplitude relation 
(\ref{ampl-rel1}), we obtain the triangle relations
\begin{eqnarray}
P+T+A&=&0\\
\overline{P}+\overline{T}+\overline{A}&=&0\,.
\end{eqnarray}
If, moreover, we neglect rescattering effects, we have $P=\overline{P}$,
and consequently arrive at the two triangles represented by the thick
solid lines in Fig.\ \ref{fig:EW-det}. If the angle $\varphi$ is known from 
the $B_s\to K^+K^-$ analysis, both $\gamma$ and $|T|$ can be simultaneously 
determined by requiring $|T|=|\overline{T}|$. Using the strategies 
proposed in \cite{defan,rf-FSI}, which make use of $B^\pm\to K^\pm K$ decays, 
rescattering processes can be taken into account in this approach to
determine $\gamma$. Its theoretical accuracy is limited by $SU(3)$-breaking 
effects and ``colour-suppressed'' electroweak penguins. Let us note that 
if $\varphi$ is unknown, $|T|$ has to be fixed in order to extract $\gamma$. 
This construction then corresponds to the one proposed in \cite{PAPIII}. 
If rescattering processes play a minor role and the hypothesis of 
``colour suppression'' works in $B\to\pi K$ decays, we have $T+C\approx T$, 
and can determine $|T|$ with the help of (\ref{T-C-det}). Moreover, if we 
use in addition the amplitudes $B\equiv\sqrt{2}\,A(B^+\to\pi^0K^+)$ and 
$\overline{B}\equiv\sqrt{2}\,A(B^-\to\pi^0K^-)$, the electroweak penguin 
amplitude $P_{\rm ew}$ can be determined \cite{PAPIII,rev}. To this end, the
relation (\ref{ampl-rel2}) with $C=0$ is used:
\begin{equation}
P+T+B+P_{\rm ew}=0\,,
\end{equation} 
as well as its CP conjugate with $P=\overline{P}$, which holds for small 
rescattering effects:
\begin{equation}
P+\overline{T}+\overline{B}+P_{\rm ew}=0\,.
\end{equation}
This strategy is also illustrated in Fig.\ \ref{fig:EW-det}, where the 
amplitudes $B$, $\overline{B}$ and $P_{\rm ew}$ are represented by the 
thin dotted lines. 

\section{Conclusions}\label{concl}
In summary, we have performed an analysis of the combinations $B^\pm\to
\pi^\pm K$, $\pi^0K^\pm$ and $B_d\to\pi^0K$, $\pi^\mp K^\pm$ of charged 
and neutral $B$ decays within a completely general formalism, taking into 
account both electroweak penguin and rescattering effects. Originally, this 
formalism was developed in \cite{defan} to probe the CKM angle $\gamma$ with 
the help of the decays $B^\pm\to\pi^\pm K$ and $B_d\to\pi^\mp K^\pm$,
but it can also be applied to these combinations of charged and neutral $B$
decays, if straightforward replacements of variables are performed. 
In this manner, we could obtain a unified picture of $B\to\pi K$ decays,
which is useful for the comparison of the various approaches using these
modes to probe the CKM angle $\gamma$.

Following these lines, we were in a position to generalize the 
strategies to constrain and determine $\gamma$ with the help of 
$B^\pm\to\pi^\pm K$, $\pi^0K^\pm$ decays, which were recently pointed out
by Neubert and Rosner \cite{nr1,nr2}. This allowed us to investigate the
sensitivity of these methods to the various assumptions made in 
\cite{nr1,nr2}, in particular to the neglect of rescattering processes of 
the kind $B^+\to\{\pi^0K^+\}\to\pi^+K^0$. We have demonstrated that 
such final-state interaction effects may lead to uncertainties similar 
to those affecting the $B^\pm\to\pi^\pm K$, $B_d\to\pi^\mp K^\pm$ strategies. 
It would be indicated experimentally that such processes play in fact an 
important role, if future experiments should find a sizeable value of the 
CP asymmetry arising in $B^\pm\to\pi^\pm K$, or a significant enhancement 
of the $B^\pm\to K^\pm K$, $B_d\to K^+K^-$ branching ratios with respect 
to their ``short-distance'' expectations. In this case, our completely
general formalism would allow us to take into account the rescattering 
effects with the help of the strategies proposed in \cite{defan,rf-FSI},
making use of $B^\pm\to K^\pm K$ decays. Unfortunately, it is not possible to
control also non-factorizable $SU(3)$-breaking effects in a similar manner.
Using our general formulae, we found that such effects may have an 
important impact on the information on the CKM angle $\gamma$ provided 
by the $B^\pm\to\pi^\pm K$, $\pi^0K^\pm$ modes.

We have also proposed a new strategy to probe the CKM angle $\gamma$ with 
the help of the neutral decays $B_d\to\pi^0K$, $\pi^\mp K^\pm$, requiring
a time-dependent analysis of $B_d\to\pi^0K_{\rm S}$. Although this method 
is more difficult from an experimental point of view, it is theoretically 
cleaner than the $B^\pm\to\pi^\pm K$, $\pi^0K^\pm$ approach. The point is
that final-state interaction effects can be taken into account in 
a clean way with the help of the direct and mixing-induced CP-violating
observables of the decay $B_d\to\pi^0K_{\rm S}$. However, the uncertainties 
related to non-factorizable $SU(3)$-breaking corrections are similar to 
those affecting the $B^\pm\to\pi^\pm K$, $\pi^0K^\pm$ strategy. 

In addition to an accurate measurement of all charged and neutral $B\to\pi K$ 
modes, an analysis of the decay $B_s\to K^+K^-$ would be very useful, allowing 
a variety of ways to combine its observables with those of the $B\to\pi K$
decays to probe the CKM angle $\gamma$ and to obtain insights into  
electroweak penguins. The former decays can already be studied 
at the $e^+$--\,$e^-$ $B$-factories (BaBar, BELLE, CLEO III), which will start 
to operate at the $\Upsilon(4\,S)$ resonance in the near future. In 
fact, the CLEO collaboration has already reported the first results for these 
modes. On the other hand, dedicated $B$-physics experiments at hadron machines 
appear to be the natural place to explore $B_s$ decays. 

Let us now critically compare the virtues and weaknesses of the various 
approaches discussed in this paper. The advantage of the $B^\pm\to
\pi^\pm K$, $\pi^0K^\pm$ and $B_d\to\pi^0K$, $\pi^\mp K^\pm$ strategies
in comparison with the one using the decays $B^\pm\to\pi^\pm K$ and 
$B_d\to\pi^\mp K^\pm$ is that the parameters $r_{({\rm c, n})}$ can
be determined with the help of the decay $B^+\to\pi^+\pi^0$ by using 
only the $SU(3)$ flavour symmetry, and that the electroweak penguins 
can be theoretically controlled by again making use of $SU(3)$ flavour 
symmetry arguments. However, a possible weakness of this approach is the
fact that non-factorizable $SU(3)$-breaking corrections, which may have 
an important impact, cannot be treated in a quantitative way at present.
In the $B^\pm\to\pi^\pm K$, $B_d\to\pi^\mp K^\pm$ strategy, ``factorization'' 
or ``colour suppression'' has to be employed to fix the parameter $r$, 
and it is more difficult to control the electroweak penguins theoretically. 
Their importance is strongly related to rescattering effects and to the 
question of ``colour suppression'' in $B\to\pi K$ decays, which can be 
probed, for instance, through the CP-violating observables of the decay 
$B_d\to\pi^0K_{\rm S}$. 

The prospects to probe $\gamma$ with the help of the decays 
$B^\pm\to\pi^\pm K$ and $B_d\to\pi^\mp K^\pm$ are good, even if the 
present central value of $R=1$ should be confirmed by future data. 
Although the combined branching ratios of these modes would imply no 
useful constraints on $\gamma$ in this case, as soon as CP violation in 
$B_d\to\pi^\mp K^\pm$ decays is observed, contours in the $\gamma$--$r$ 
plane can be fixed, allowing the extraction of $\gamma$. In our illustrative 
example, we found contours with the interesting feature that the extracted 
value of $\gamma$ is very insensitive to the value of $r$. Consequently, 
in such a fortunate situation, this strategy to determine $\gamma$ 
would not be weakened by the fact that the uncertainty of $r$ may be 
larger than that of $r_{\rm c}$. On the other hand, in the $B^\pm\to\pi^\pm K$,
$\pi^0 K^\pm$ case, our quantitative example gave less promising contours 
in the $\gamma$--$r_{\rm c}$ plane, where the extracted value of $\gamma$ 
shows a sizeable dependence on $r_{\rm c}$. Clearly time will tell which 
approach is more promising for the future $B$-factories.

An accurate measurement of $B$-meson decays into $\pi K$, $\pi\pi$ and
$KK$ final states is an important goal for future dedicated $B$-physics 
experiments. The physics potential of these modes is very rich, allowing 
several strategies to probe CKM phases and to shed light on the issue of 
rescattering effects and electroweak penguins. Also certain $B_s$ decays
are very useful in this respect. We are optimistic that the $B$-factory
era, which is just ahead of us, will lead to many interesting and
exciting results.

\vspace{0.5truecm}
\noindent
{\it Acknowledgements}
 
\vspace{0.3truecm}
 
\noindent
We would like to thank Gerhard Buchalla for discussions. A.J.B. would like 
to thank the CERN theory group for its hospitality during his stay at CERN. 
This work was partly supported by the German Bundesministerium f\"ur Bildung 
und Forschung under contract 06 TM 874 and by the DFG project Li 519/2-2.

\newpage

\boldmath
\section*{Appendix: Controlling Electroweak Penguins in the $\alpha$
Determination from $B\to\pi\pi$ Decays}
\unboldmath
\begin{figure}
\centerline{
\rotate[r]{
\epsfxsize=6.5truecm
\epsffile{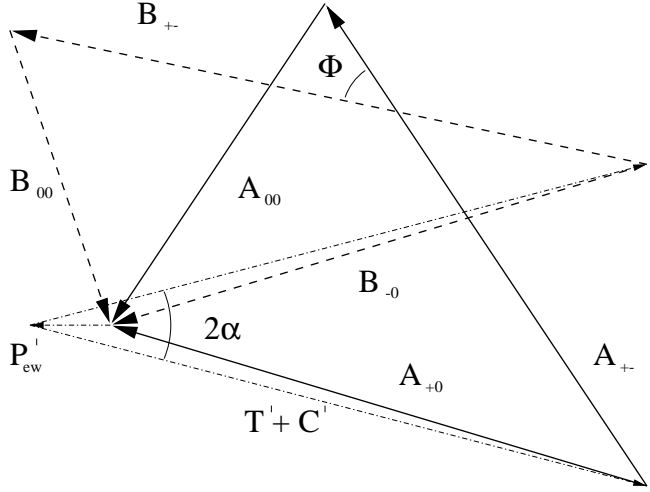}}}
\vspace*{0.7truecm} 
\caption{Determination of the CKM angle $\alpha$ by means of $B\to\pi\pi$
isospin relations in the presence of electroweak 
penguins.}\label{fig:alpha-det}
\end{figure}
In this appendix, we point out that a theoretical input similar to 
(\ref{bdEWP}) allows us to take into account electroweak penguin 
topologies in the determination of the CKM angle $\alpha$ with the 
help of the Gronau--London method \cite{gl}, using the $B\to\pi\pi$ 
isospin relation
\begin{equation}
A_{+-}+A_{00}=A_{+0}
\end{equation}
with 
\begin{equation}
A_{+-}\equiv A(B^0_d\to\pi^+\pi^-),\quad
A_{00}\equiv \sqrt{2}\,A(B^0_d\to\pi^0\pi^0),\quad
A_{+0}\equiv \sqrt{2}\,A(B^+\to\pi^+\pi^0).
\end{equation}
We have illustrated this approach in Fig.\ \ref{fig:alpha-det}, where 
\begin{equation}
B_{+-}\equiv e^{-i2\beta}\,\overline{A}_{+-},\quad
B_{00}\equiv e^{-i2\beta}\,\overline{A}_{00},\quad
B_{-0}\equiv e^{-i2\beta}\,\overline{A}_{+0},
\end{equation}
and $\beta=180^\circ-\alpha-\gamma$ denotes another angle of the unitarity 
triangle. The amplitude $A_{+0}$ can be decomposed as follows:
\begin{equation}\label{Ap0rel}
A_{+0}=-\left[\,(T'+C')+P_{\rm ew}'\,\right],
\end{equation}
where the $\bar b\to\bar d$ amplitudes $T'+C'$ and $P_{\rm ew}'$ are
defined to be proportional to the CKM factors $\lambda_u^{(d)}$ and 
$\lambda_t^{(d)}$, respectively. This definition of $P_{\rm ew}'$ is
useful in the present case, as it gives \cite{PAPI}
\begin{equation}\label{EE96}
\overline{P_{\rm ew}'}=e^{2i\beta}\,P_{\rm ew}'\,.
\end{equation}
Note that the amplitudes $(T+C)_{\bar b\to\bar d}$ and 
$(P_{\rm ew})_{\bar b\to\bar d}$ in (\ref{bdEWP}) are defined to be
proportional to $\lambda_u^{(d)}$ and $\lambda_c^{(d)}$, respectively,
which is the appropriate definition for the strategies to probe the CKM
angle $\gamma$ discussed in this paper. Proceeding as in 
Subsection~\ref{SubSec22} and using a similar theoretical input, we obtain 
(see also \cite{PAPIII})
\begin{equation}\label{EW-GL}
\left(\frac{P_{\rm ew}'}{T'+C'}\right)_{\bar b\to\bar d}=\frac{3}{2}
\left[\frac{C_9(\mu)+C_{10}(\mu)}{C_1(\mu)+C_2(\mu)}\right]
\frac{|V_{td}|}{|V_{ub}|}\,e^{i\alpha}=-\,1.3\times 10^{-2}\times
\frac{|V_{td}|}{|V_{ub}|}\,e^{i\alpha}.
\end{equation}
In contrast to (\ref{EE3}), the $SU(2)$ isospin symmetry suffices to 
derive this expression, i.e.\ no $SU(3)$ flavour symmetry arguments have
to be used to this end.

With all this information at hand, the determination of $\alpha$ from
$B\to\pi\pi$ decays in the presence of electroweak penguins can be 
accomplished as follows:
\begin{enumerate}
\item The two triangles represented by the thick solid and dashed lines can 
be determined by measuring all $B,\overline{B}\to\pi\pi$ branching 
ratios, while their relative orientation, i.e.\ the angle $\Phi$, can be 
fixed by measuring mixing-induced CP violation in the mode $B_d\to\pi^+\pi^-$ 
(a detailed discussion can be found in \cite{PAPI}). 
\item The two squashed triangles in Fig.\ \ref{fig:alpha-det} represent
the relation (\ref{Ap0rel}) and its CP conjugate, multiplied by 
$e^{-i2\beta}$. The inspection of these triangles, together with 
the phase in (\ref{EW-GL}), tells us that $P_{\rm ew}'$ lies 
on the line that bisects the angle between the amplitudes $A_{+0}$ 
and $B_{-0}$.
\item Since (\ref{EW-GL}) implies $|P_{\rm ew}'|\ll |T'+C'|$, we have, to
a very good approximation:
\begin{equation}\label{PewPr}
|P_{\rm ew}'|=1.3\times10^{-2}\times\frac{|V_{td}|}{|V_{ub}|}\,|A_{+0}|,
\end{equation}
where $|A_{+0}|$ is obtained from BR$(B^+\to\pi^+\pi^0)$. Equation 
(\ref{PewPr}), in combination with the minus sign in (\ref{EW-GL})
and the two previous steps, allows us to complete the construction shown
in Fig.\ \ref{fig:alpha-det}, and to determine the CKM angle $\alpha$. 
\end{enumerate}


\begin{thebibliography}{99}

\bibitem{cleo}CLEO Collaboration (R. Godang et al.), 
{\it Phys.\ Rev.\ Lett.}~{\bf 80} (1998) 3456.

\bibitem{ICHEP98}For a review, see R. Fleischer, preprint CERN-TH/98-296 
(1998) [hep-ph/9809302], invited talk given at the 29th International 
Conference on High Energy Physics (ICHEP '98), Vancouver, Canada, 23--29 
July 1998; to appear in the proceedings.

\bibitem{PAPIII}R. Fleischer, {\it Phys.\ Lett.}~{\bf B365} (1996) 399.

\bibitem{groro}M. Gronau and J.L. Rosner, {\it Phys.\ Rev.}~{\bf D57} (1998)
6843.
 
\bibitem{fm2}R. Fleischer and T. Mannel, {\it Phys.\ Rev.}~{\bf D57}
(1998) 2752.

\bibitem{defan}R. Fleischer, {\it Eur.\ Phys.\ J.}~{\bf C} (1998) 
DOI 10.1007/s100529800919 [hep-ph/9802433]. 

\bibitem{rf-FSI}R. Fleischer, {\it Phys. Lett.}~{\bf B435} (1998) 221.
 
\bibitem{newCLEO}CLEO Collaboration (M. Artuso et al.), preprint
CLEO CONF 98-20; J. Alexander, plenary talk given at the 29th International 
Conference on High Energy Physics (ICHEP '98), Vancouver, Canada, 23--29 
July 1998; to appear in the proceedings.

\bibitem{gnps}Y. Grossman, Y. Nir, S. Plaszczynski and M.-H. Schune, 
{\it Nucl.\ Phys.}~{\bf B511} (1998) 69.

\bibitem{grl}M. Gronau, J.L. Rosner and D. London, {\it Phys.\ Rev.\ 
Lett.}~{\bf 73} (1994) 21. 

\bibitem{hlgr}O.F. Hern\'andez, D. London, M. Gronau and J.L. Rosner,
{\it Phys.\ Lett.}~{\bf B333} (1994) 500; {\it Phys.\ Rev.}~{\bf D50} 
(1994) 4529.

\bibitem{dh}N.G. Deshpande and X.-G. He, {\it Phys.\ Rev.\ 
Lett.}~{\bf 74} (1995) 26 [E: {\bf 74} (1995) 4099]. 

\bibitem{rf-ewp}R. Fleischer, {\it Z. Phys.}~{\bf C62} (1994) 81; 
{\it Phys.\ Lett.}~{\bf B321} (1994) 259.

\bibitem{rf-ewp3}R. Fleischer, {\it Phys.\ Lett.}~{\bf B332} (1994) 419.

\bibitem{akl}For a recent study, see A. Ali, G. Kramer and C.-D. L\"u, 
{\it Phys.\ Rev.}~{\bf D58} (1998) 094009.

\bibitem{neubert}M. Neubert, {\it Phys.\ Lett.}~{\bf B424} (1998) 152.

\bibitem{PAPI}A.J. Buras and R. Fleischer, {\it Phys.\ Lett.}~{\bf B365} 
(1996) 390.

\bibitem{nr1}M. Neubert and J.L. Rosner, preprint CERN-TH/98-273 (1998)
[hep-ph/9808493].

\bibitem{nr2}M. Neubert and J.L. Rosner, preprint CERN-TH/98-293 (1998)
[hep-ph/9809311].

\bibitem{FSI}L. Wolfenstein, {\it Phys.\ Rev.}~{\bf D52} (1995) 537; 
J. Donoghue, E. Golowich, A.~Petrov and J. Soares, {\it Phys.\ Rev.\ 
Lett.}~{\bf 77} (1996) 2178; B. Blok and I. Halperin, {\it Phys.\ 
Lett.}~{\bf B385} (1996) 324; B. Blok, M. Gronau and J.L.~Rosner,
{\it Phys.\ Rev.\ Lett.}~{\bf 78} (1997) 3999.

\bibitem{gewe}J.-M. G\'erard and J. Weyers, preprint UCL-IPT-97-18 (1997) 
[hep-ph/9711469].

\bibitem{fknp}A.F. Falk, A.L. Kagan, Y. Nir and A.A. Petrov, 
{\it Phys.\ Rev.}~{\bf D57} (1998) 4290.

\bibitem{atso}D. Atwood and A. Soni, {\it Phys.\ Rev.}~{\bf D58} (1998) 
036005. 

\bibitem{groro-FSI}M. Gronau and J.L. Rosner, preprint EFI-98-23 (1998) 
[hep-ph/9806348].

\bibitem{bfm}A.J. Buras, R. Fleischer and T. Mannel, preprint
CERN-TH/97-307 (1997) [hep-ph/9711262], to appear in 
{\it Nucl.\ Phys.}~{\bf B}.

\bibitem{wolf}L. Wolfenstein, {\it Phys.\ Rev.\ Lett.}~{\bf 51} (1983) 1945. 

\bibitem{blo}A.J. Buras, M.E. Lautenbacher and G. Ostermaier, {\it Phys.\
Rev.}~{\bf D50} (1994) 3433.

\bibitem{Rb-update}P. Rosnet, talk given at the 29th International 
Conference on High Energy Physics (ICHEP '98), Vancouver, Canada, 23--29 
July 1998; to appear in the proceedings.

\bibitem{BBL}G. Buchalla, A.J. Buras and M.E. Lautenbacher, {\it Rev.\
Mod.\ Phys.}~{\bf 68} (1996) 1125.

\bibitem{nq}Y. Nir and H.R. Quinn, {\it Phys.\ Rev.\ Lett.}~{\bf 67} 
(1991) 541.

\bibitem{ghlr}M. Gronau, O.F. Hern\'andez, D. London and J.L. Rosner, 
{\it Phys.\ Rev.}~{\bf D52} (1995) 6374.

\bibitem{groro-rel}M. Gronau and J.L. Rosner, preprint SLAC-PUB-7945 (1998) 
[hep-ph/9809384].

\bibitem{wuegai}F. W\"urthwein and P. Gaidarev, preprint
CALT-68-2153 (1997) [hep-ph/9712531].

\bibitem{UTfits}This range corresponds to an update of the analysis
performed in: A.J. Buras, preprint TUM-HEP-316-98 (1998) [hep-ph/9806471];
to appear in {\it Probing the Standard Model of Particle Interactions}, 
eds.\ F. David and R. Gupta (Elsevier Science B.V., Amsterdam, 1998). 

\bibitem{gl}M. Gronau and D. London, {\it Phys.\ Rev.\ Lett.}~{\bf 65} 
(1990) 3381.

\bibitem{rev}R. Fleischer, {\it Int.\ J. Mod.\ Phys.}~{\bf A12} (1997) 2459.

\bibitem{csbs}A.B. Carter and A.I. Sanda, {\it Phys.\ Rev.\ Lett.}~{\bf 45}
(1980) 952; {\it Phys.\ Rev.}~{\bf D23} (1981) 1567; I.I. Bigi and A.I. 
Sanda, {\it Nucl.\ Phys.}~{\bf B193} (1981) 85.

\bibitem{PAPII}A.J. Buras and R. Fleischer, {\it Phys.\ Lett.}~{\bf B360} 
(1995) 138.

\bibitem{bsgam}R. Fleischer, {\it Phys.\ Rev.}~{\bf D58} (1998) 093001.

\bibitem{fd1}R. Fleischer and I. Dunietz, {\it Phys.\ Rev.}~{\bf D55} (1997)
259.

\bibitem{kly}C.S. Kim, D. London and T. Yoshikawa, {\it Phys.\ Rev.}~{\bf D57}
(1998) 4010.

\bibitem{DGamma}For a recent calculation of $\Delta\Gamma_s$, see 
M. Beneke, G. Buchalla, C. Greub, A. Lenz and U. Nierste, preprint 
CERN-TH/98-261 (1998) [hep-ph/9808385].
 
\bibitem{dunietz}I. Dunietz, {\it Phys.\ Rev.}~{\bf D52} (1995) 3048.

\bibitem{ddf1}A.S. Dighe, I. Dunietz and R. Fleischer, [hep-ph/9804253],
{\it Eur.\ Phys.\ J.}~{\bf C} (1998) DOI 10.1007/s100529800954.

\bibitem{ghlr2}M. Gronau, O.F. Hern\'andez, D. London and 
J.L. Rosner, {\it Phys.\ Rev.}~{\bf D52} (1995) 6356.

\bibitem{adk}R. Aleksan, I. Dunietz and B. Kayser, {\it Z. Phys.}~{\bf C54}
(1992) 653.
 
\end{thebibliography}
\end{document}